\documentclass[sigconf]{acmart}

\usepackage{booktabs}

\usepackage[normalem]{ulem}
\usepackage{subfigure}

\usepackage{color, colortbl}

\setcopyright{acmcopyright}
\copyrightyear{2022}
\acmYear{2022}
\acmDOI{XXXXXXX.XXXXXXX}

\acmConference[SIGSPATIAL '22]{ACM SIGSPATIAL International Conference on Advances in Geographic Information Systems}{November 01--04,
  2022}{Seattle, WA}
\acmPrice{15.00}
\acmISBN{978-1-4503-XXXX-X/18/06}

\begin{document}

\title{How Routing Strategies Impact Urban Emissions}

\author{Giuliano Cornacchia}
\affiliation{%
  \institution{University of Pisa, ISTI-CNR}
  \city{Pisa}
  \country{Italy}
}
\email{giuliano.cornacchia@phd.unipi.it}

\author{Matteo B\"{o}hm}
\affiliation{%
  \institution{Sapienza University of Rome}
  \city{Rome}
  \country{Italy}
}
\email{bohm@diag.uniroma1.it}

\author{Giovanni Mauro}
\affiliation{
  \institution{University of Pisa, IMT, ISTI-CNR}
  \city{Pisa}
  \country{Italy}
}
\email{giovanni.mauro@phd.unipi.it}

%

\author{Mirco Nanni}
\affiliation{%
 \institution{ISTI-CNR}
 \streetaddress{Via G. Moruzzi 1}
 \city{Pisa}
 \country{Italy}}
 \email{mirco.nanni@isti.cnr.it}

\author{Dino Pedreschi}
\affiliation{%
  \institution{University of Pisa}
  \streetaddress{Largo Bruno Pontecorvo}
  \city{Pisa}
  \country{Italy}}
  \email{dino.pedreschi@unipi.it}

\author{Luca Pappalardo}
\affiliation{%
  \institution{ISTI-CNR}
  \streetaddress{Via G. Moruzzi 1}
  \city{Pisa}
  \country{Italy}
  \postcode{56123}}
\email{luca.pappalardo@isti.cnr.it}

\renewcommand{\shortauthors}{Cornacchia et al.}

\begin{abstract}
Navigation apps use routing algorithms to suggest the best path to reach a user's desired destination. Although undoubtedly useful, navigation apps' impact on the urban environment (e.g., carbon dioxide emissions and population exposure to pollution) is still largely unclear. In this work, we design a simulation framework to assess the impact of routing algorithms on carbon dioxide emissions within an urban environment. Using APIs from TomTom and OpenStreetMap, we find that settings in which either all vehicles or none of them follow a navigation app's suggestion lead to the worst impact in terms of CO$_2$ emissions. In contrast, when just a portion (around half) of vehicles follow these suggestions, and some degree of randomness is added to the remaining vehicles' paths, we observe a reduction in the overall CO$_2$ emissions over the road network. Our work is a first step towards designing next-generation routing principles that may increase urban well-being while satisfying individual needs. 
\end{abstract}



\maketitle

\section{Introduction}
\label{sec:introduction}

To tackle the pressing challenge of climate change \cite{bai2018six} and the urgent call by the United Nations to reduce the adverse environmental impact of cities \cite{assembly2015sustainable}, there is an increasing effort to study the impact of human mobility on urban well-being dimensions, such as traffic congestion, air pollution, and carbon dioxide emissions.
An element that enriches the complexity of this challenge is the widespread diffusion of GPS navigation apps such as TomTom, Google Maps, and Waze, which use routing algorithms, heuristics and AI to suggest the best path to reach a user's desired destination. 
Although undoubtedly useful, particularly when exploring an unfamiliar city,
GPS navigation apps may also cause several issues in the urban environment: since they are typically optimised to keep an individual's trip as short as possible, they do not care about collective effects on the city, such as whether the traffic can be absorbed by the streets, compromises safety or creates more pollution \cite{debaets2014route, macfarlane2019when, siuhi2016opportunities}. 
Many documented cases show that GPS navigation apps may create chaos: for example, they sometimes divert heavy traffic through side roads in nearby towns, with such an impact on the residents' life that experts stated that these apps are a new agent claiming a ``right to the city"~\cite{leonia, fisher2022algorithms}.

Beyond the anecdotal, preliminary research show that the impact of navigation apps on the urban environment is mixed \cite{ericsson2006optimizing, samaras2016quantification}. 
On the one hand, navigation apps may provide benefits in mitigating carbon dioxide emissions \cite{arora2021quantifying}; on the other hand, they may increase the population exposure to pollution in densely-populated areas \cite{perezprada2017managing}. 
Overall, existing studies are sporadic and yield contradictory results, leading to a picture of the navigation apps' impact on the urban environment that is mainly unclear and incomplete. 
In particular, the literature still lacks a rigorous framework to assess and compare the impact of navigation apps on urban well-being.

In this paper, we design a simulation framework -- TraffiCO$_2$ -- to assess the impact of GPS navigation apps on urban well-being in terms of carbon dioxide (CO$_2$) emissions.
The framework uses GPS data to reconstruct a city's mobility demand, and navigation apps' APIs to obtain routing suggestions (paths on the road network) for a set of vehicles' origin and destination.
Then, TraffiCO$_2$ relies on a traffic simulator that considers all aspects of vehicular mobility (e.g., jams, queues at traffic lights, roads capacity), to generate trajectories that describe when each vehicle visits each road in its path.
Finally, the framework use an emission model to estimate CO$_2$ emissions for each segment in the road network.

We apply TraffiCO$_2$ to the city of Milan (Italy) to study how emissions changes as we increase the fraction of vehicles that follow navigation apps' routing.
We find that settings in which either all vehicles or none of them follow a navigation apps' routing suggestion lead to the highest amount of CO$_2$ emissions. 
These settings also correspond to the most uneven distribution of the CO$_2$ emissions across the roads, with a few roads suffering the greater quantity of emissions.
In contrast, a scenario where just a fraction of the vehicles (around 50\%) follows navigation apps' suggestion and the remaining part follow a randomly perturbed fastest path, reduces the vehicles' overall emissions and distributes them more evenly on the road network. 
We also find that the navigation apps' routing affects the spatial distribution of emissions, making Milan's external ring road more polluted while unloading the internal roads from the emissions.
Notably, adding perturbation to the vehicles' path is beneficial, as it reduces the overall emissions, distributes them more evenly, and reduces the vehicles' average travel time.

Our simulation framework is a useful tool to assess and compare routing strategies, helping drivers, institutions, and policymakers understand the impact of navigation apps on the urban environment. 
Moreover, TraffiCO$_2$ is a first step towards designing and testing next-generation routing principles that may increase urban well-being while satisfying individual needs.

\subsection*{Open Source}
We provide the implementation of TraffiCO$_2$, the code and the link to the data to reproduce our study at \url{https://bit.ly/traffico2_gh}.

\section{Related Work}
\subsection*{Environmental impact of vehicular traffic}
The environmental impact generated by vehicles (e.g., air pollution) is becoming increasingly evident in urban environments.
Existing methods to quantify vehicles' emissions range between two extremes. On the one hand, some approaches rely on measurements performed on small samples of vehicles with high spatio-temporal resolutions, such as those coming from particulate sensors~\cite{desouza2020} or portable emissions measurement systems (PEMS)~\cite{chong2020,lujan2018}. 
These sensors measure emissions in real-world driving conditions, producing accurate estimates but hardly generalisable patterns due to the limited sample size. For example, two studies~\cite{lujan2018,chong2020} analyse emissions from PEMS of one and three vehicles, finding that the highest emissions are associated with the urban part of the route, flat roads, and low speed.

On the other hand, some studies cover a region's entire fleet, such as those using odometer readings from annual safety inspections.
These data describe each vehicle's age, fuel type, engine volume, and mileage, used in macroscopic models to estimate annual emissions.
Two studies~\cite{chatterton2015use,diao2014vehicle} use odometer readings to compute mean annual emissions for UK postcode areas and explore the built-environment effects (e.g., work accessibility) on the vehicles' annual miles travelled in Boston.
Unfortunately, odometer readings miss critical information such as instantaneous speed and acceleration~\cite{kancharla2018incorporating,choudhary2016urban,ferreira2015impact,zheng2017influence}, making it challenging to track emissions over time and map them to suburban areas.

Somewhere between these two extremes lie works that use GPS traces, which describe human mobility in great detail~\cite{luca2020deep,barbosa2018human} and can cover a representative fraction of the vehicle fleet~\cite{pappalardo2013understanding, bohm2021improving}. 
These data allow computing instantaneous speed and acceleration, which can be used within microscopic models to obtain emissions estimates in high spatio-temporal resolution.
Several studies use GPS traces to analyse the vehicles' emissions at different spatio-temporal scales~\cite{bohm2021improving,nyhan2016,liu2019}, investigate the relationship between emissions and the urban environment~\cite{reznik2018}, vehicle miles travelled and fuel consumption~\cite{wang2014using}, or trip rates and travel mode choice~\cite{cervero1997travel}. 
Other studies concentrate on congestion-related emissions~\cite{gately2017urban} or braking~\cite{chen2020mining}, emissions associated with ride-hailing~\cite{sui2019gps} and bus stops' positioning~\cite{yu2020mobile}, the impact of urban policies~\cite{rahman2017tribute}, methods for emission modelling~\cite{zhu2020high,aziz2018novel}, and air quality monitoring~\cite{desouza2020}.

\subsection*{Microscopic traffic simulation} 
Microscopic traffic simulators are crucial to simulate vehicular traffic given a mobility demand.
A notable example is SUMO (Simulation of Urban MObility), an open-source, portable, and multi-modal tool designed to handle traffic simulations on road networks~\cite{Microscopic2018Lopez, alazzawi2018simulating, krajzewicz2012recent}. 
SUMO allows controlling several aspects of traffic, from fuel consumption to vehicle emissions and routing strategies.
Several works use SUMO to study the impact of vehicular traffic on the urban environment.
Alazzawi et al.~\cite{alazzawi2018simulating} simulate the introduction of a fleet of automated shared vehicles into Milan, finding that a fleet of 9500 of these vehicles helps mitigate traffic congestion and emissions.
Malik et al.~\cite{malik2019evaluation} propose a traffic system that re-routes an emergency vehicle in case of traffic jams to reduce travel time and pollution. 
Krajzewicz et al.~\cite{krajzewicz2005simulation} use optical information systems to optimise traffic lights over junctions better than traditional approaches.
Zubillaga et al.~\cite{zubillaga2014measuring} compare the traditional traffic-light coordination (green-wave method) with a self-organising method that adapts to traffic demands, showing how the latter is way better. 

\subsection*{Urban routing and impact of navigation systems}
People's natural routing choices may significantly deviate from the optimal route.
The origin of these sub-optimal human routes may lie in several factors, e.g., the environment in which one grew up \cite{coutrot2022entropy}, the subjective perception of space \cite{norman2005perception}, the presence of landmarks \cite{foo2005do}, and even the usage of an electric vehicle \cite{jensen2020route}.

Zhu et al.~\cite{zhu2015do} find that only $34\%$ of trips, mostly very short and very long ones, follow the shortest time path.
Lima et al.~\cite{lima2016understanding} find that $53\%$ of the drivers' routes are not optimal and that most drivers use a few preferred routes for their journeys.
Xu et al.~\cite{xu2021understanding} find similar results using location-based services data, which are less accurate but more pervasive than vehicular trajectory data.
Bongiorno et al.~\cite{bongiorno2021vector} find that people increasingly deviate from the shortest path as the distance between origin and destination increases. 
Similarly, Manley et al.~\cite{manley2015shortest} show that people's route choice results from multiple decisions made at each anchor of the route.
Given the evident uncertainty in people's routing strategies, the impact of navigation apps' routing suggestions on the urban environment is not negligible~\cite{macfarlane2019when}.
For this reason, some effort has been put into developing ``environmentally-friendly'' navigation apps that minimise fuel consumption instead of travel time~\cite{barth2007environmentally}.
Studies on the effects of this eco-routing, either as fuel savings~\cite{ericsson2006optimizing} or system-wide impact~\cite{ahn2013network,samaras2016quantification}, find that the urban impact of navigation apps is mixed: green navigation apps can reduce CO$_2$ emissions but increase the population exposure to nitrogen oxides~\cite{perezprada2017managing}. 
Following the same research line, Mehrvarz et al.~\cite{mehrvarz2020optimal} find that the fastest route suggested by navigation apps may not optimise fuel consumption.
In contrast, Arora et al.~\cite{arora2021quantifying} show that following Google Maps' routing strategy saves $1.7\%$ of CO$_2$ emissions and $6.5\%$ travel time. 

\subsection*{Position of our work}
Existing studies are sporadic and yield contradictory results, and the impact of navigation apps on urban well-being is mainly unclear and incomplete.
Moreover, the existing literature lacks a rigorous framework to assess and compare the various navigation apps and routing strategies.
We fill this gap by proposing TraffiCO$_2$, a simulation framework to estimate the adverse impact of different routing strategies on total CO$_2$ emissions and their distribution of a city's road network.
Thus, our work extends the branch of literature on sustainable urban mobility by providing a rigorous experimental framework to disentangle the impact of routing criteria on the urban environment. 

\begin{figure*}[htb!]
    \centering
    \includegraphics[width=\linewidth]{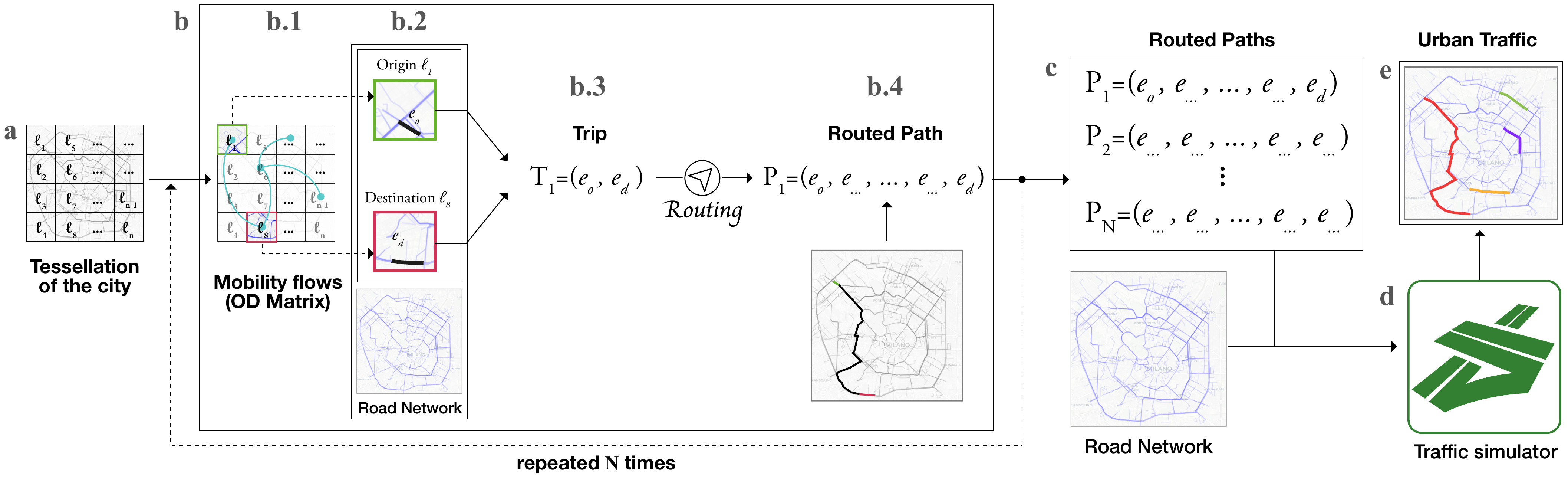}
    \caption{
Schema of the TraffiCO$_2$ simulation framework.
(a) The city is split into squared tiles using scikit-mobility \cite{scikitmob}. 
(b.1) Real data are used to estimate mobility flows (OD matrix) within the city. (b.2-b.3) A trip is created by selecting at random an origin-destination pair from the OD matrix and two edges on the road network.
(b.5) Some routing algorithm is used to convert each trip into a path on the road network. 
(c) Steps b.1-b.5 are repeated $N$ times ($N$ = number of vehicles) to obtain a multiset of routed paths.
(d) A traffic simulator (SUMO) is used to simulate the urban traffic generated by the routed paths (e).}
    \label{fig:schema}
\end{figure*}

\section{Simulation Framework}
\label{sec:simulation_framework}
We design a simulation framework -- TraffiCO$_2$ -- to generate a realistic urban traffic considering different routing strategies (Figure~\ref{fig:schema}).
First, we use real data to generate a \textit{mobility demand} describing trips (origin-destination pairs) in an urban environment.
Second, we transform each trip into a path on the road network using some routing algorithm, obtaining a multiset of \textit{routed paths}.
Third, we use an agent-based model (SUMO) that, considering realistic aspects of vehicular mobility (e.g., jams, traffic lights, slowdowns), simulates an \textit{urban traffic} based on the multiset of routed paths.
Finally, we compute the vehicles' travel time and the CO$_2$ emissions on each road from the trajectories generated by SUMO.

\subsection{Road Network} 
\label{sec:road_network}
It describes the road infrastructure where the vehicles move during the simulation. 
The road network is a directed graph $G=(V, E)$, where $V$ is the set of nodes representing road intersections and $E$ is the set of edges representing roads. 
Both nodes and edges may have attributes, such as: traffic lights, number of lanes, road speed limit and type (e.g., motorway, secondary road).
These attributes are used by SUMO to simulate realistic aspects of vehicular mobility.

\subsection{Mobility Demand}
\label{sec:mobility_demand}
The \emph{mobility demand} $D = \{T_1, \dots, T_N\}$ is a multiset of $N$ trips (one per each vehicle) within an urban environment. 
A single trip $T_v=(o,d)$ for a vehicle $v$ is defined by its origin location $o$ and destination location $d$.
To compute $D$, we first divide the urban environment into squared tiles of a given side (Figure \ref{fig:schema}a). Second, we use real mobility data to estimate the flows between the tiles, thus obtaining an origin-destination matrix $M$ where an element $m_{o, d}\in M$ describes the number of vehicles' trips that start in tile $o$ and end in tile $d$ (Figure \ref{fig:schema}b.1). 

Then, we iterate $N$ times the following procedure. A vehicle's $v$ trip is a pair $T_v=(e_o, e_d)$ generated by selecting at random a matrix element $m_{o,d} \in M$ with a probability $p_{o, d} \propto m_{o, d}$ and uniformly at random two edges $e_o, e_d \in E$ within tiles $o$ and $d$, respectively (Figure \ref{fig:schema}b.2, b.3). 

\subsection{Paths generation}
\label{sec:paths_generation}
We translate the $N$ trips in $D$ into $N$ paths obtaining a multiset $\overline{D}$ of \emph{routed paths} within the urban environment. 
Each path $P_v(e_o,e_d, R) {=} (e_o, \dots, e_d)$ of a vehicle $v$
is a sequence of edges on the road network connecting $e_o$ and $e_d$ (Figure \ref{fig:schema}b.5), obtained by some routing algorithm $R$. 
When a vehicle's path is generated by a routing algorithm $R$, we say that the vehicle is $R$-routed.
In $\overline{D} = {\{P_1, \dots, P_N\}}$, the routed paths (Figure \ref{fig:schema}c) are generated independently by (different) routing algorithms. 

\subsection{Traffic Simulation}
\label{sec:traffic_simulation}
We simulate the vehicular traffic generated by the routed paths in $\overline{D}$ using SUMO (Simulation of Urban MObility) \cite{Microscopic2018Lopez, alazzawi2018simulating, krajzewicz2012recent} (Figure~\ref{fig:schema}d). 
SUMO explicitly models each vehicle's physics and dynamics, including their routes through the road network, allowing us to simulate vehicular traffic realistically, including traffic jams, queues at traffic lights, and slowdowns due to heavy traffic.
SUMO outputs an \emph{urban traffic} (Figure \ref{fig:schema}e), i.e., a multiset $S(\overline{D}, \mathcal{T}) = \{\overline{P_1}, \dots, \overline{P_N}\}$ where: 
\begin{itemize}
\item $\mathcal{T} = (t^{(1)}, \dots, t^{(N)})$ is the sequence of departure times of the $N$ paths in $\overline{D}$, where $t^{(i)} \in \mathcal{T}$ is a timestamp chosen uniformly at random from the simulation interval  (e.g., 1 hour); 
\item a vehicle $v$'s trajectory $\overline{P_v} \in S(\overline{D},\mathcal{T})$ is defined as: $$\overline{P_v} = ((e_o, t_1^{(v)}), \dots, (e_d, t_m^{(v)}))$$ where $m$ is the length of path $\overline{P_v}$.

\end{itemize}

\subsection{Vehicle Emissions}
\label{sec:vehicle_emissions}
We assess the impact of the urban traffic $S(\overline{D}, \mathcal{T})$ using a model that estimates the vehicles' emissions based on their trajectories $\overline{P_1}, \dots, \overline{P_N}$. 
Specifically, we use the HBEFA3 emission model based on the Handbook of Emission Factors for Road Transport (HBEFA) database~\cite{infras2013handbook}. 
The HBEFA3-based model estimates the vehicle's instantaneous CO$_2$ emissions relying on the following function, which is linked to the power the vehicle's engine produces in each trajectory point $j$ to overcome the driving resistance force~\cite{krajzewicz2015second}:
$$\mathcal{E}(j) = c_0 + c_1sa + c_2sa^2 + c_3s + c_4s^2 + c_5s^3$$
where $s$ and $a$ are the  vehicle's speed and acceleration in point $j$, respectively, and $c_0,\dots,c_5$ are parameters changing
per emission type and vehicle taken from the HBEFA database.

We compute the amount of CO$_2$ emissions on each edge $e \in E$ by summing all the emissions corresponding to any vehicle $v$'s trajectory point that fall on $e$, i.e., $\mathcal{E}(e) = \sum_v \sum_{j \in \overline{P_v}} \mathcal{E}(j)$.
Finally, we construct a weighted road network $\overline{G} = (V, \overline{E})$ where each edge $\overline{e} \in \overline{E}$ is associated with the attribute $\mathcal{E}(e)$ describing the amount of CO$_2$ emissions on it.

\section{Experimental Setup}
\label{sec:experimental_setup}
We apply TraffiCO$_2$ to a 45 km$^2$ area in the city centre of Milan, Italy, for which we have GPS data describing 17k private vehicles travelling between April 2nd and 8th, 2007 (114k GPS points).
We split Milan into squared tiles with 1 km side, detect the tiles where each vehicle starts and stops \cite{Ramaswamy2004Project, scikitmob}, and compute the origin-destination matrix $M$ of vehicles' flows.

We download Milan's road network $G=(V, E)$ from OpenStreetMap ($|V|=5551$ nodes and $|E|=36,945$ edges) and preprocess it to fix incorrect information regarding turns, intersections, road interruptions, the number of lanes per road, and other inaccuracies that characterise these data \cite{Argota2022Getting}. 

Based on the matrix $M$ and the road network $G$, we compute the mobility demand $D$ with $N=15,000$ vehicles using the procedure described in Section \ref{sec:mobility_demand}.
We choose $N=15,000$ because it minimises the difference between the distribution of travel time of real trajectories and simulated ones, a common way to assess the realism of a simulated urban traffic \cite{Argota2022Getting} (see Appendix \ref{sec:traf-cal} for detail). 

To translate the mobility demand $D$ into the multiset of routed paths $\overline{D}$, we consider two routing algorithms: OpenStreetMap (OSM) and TomTom (TT). 
OSM is a public voluntary geographic information system, which provides APIs\footnote{\href{https://openrouteservice.org/dev/\#/api-docs}{https://openrouteservice.org/dev/\#/api-docs}} to generate paths between locations.
TT is a commercial navigation system service that provides APIs\footnote{\href{https://developer.tomtom.com/routing-api/documentation/product-information/introduction}{https://developer.tomtom.com/routing-api}} for routes generation.
For OSM and TT, we use a routing principle that suggests a path between two locations as a trade-off between travel time and distance.
Note that OSM and TT implement this principle differently, i.e., they may provide different paths for the same origin-destination pair.

We obtain the path of vehicles that do not follow navigation apps' suggestions using Duarouter (DR), a routing algorithm provided by SUMO\footnote{\href{https://sumo.dlr.de/docs/duarouter.html}{https://sumo.dlr.de/docs/duarouter.html}} that suggests the fastest path (i.e., shortest travel time) between two edges on the road network.
The fastest path may be perturbed using a randomisation parameter $w \in [1, +\infty)$, where $w=1$ means no randomisation (i.e., the fastest path). The higher $w$, the more randomly perturbed the fastest path is.
Therefore, DR with $w > 1$ allows us to model the driving behaviour of vehicles that do not strictly follow navigation apps' suggestions and, simultaneously, to model the imperfection and non-rationality of human drivers \cite{Seele2012Cognitive}. 
Indeed, individuals get distracted when driving (e.g., take wrong turns), and they lack complete knowledge of the city's traffic, the road network, and the best path to reach a destination. 
In our experiments, we use $w=5$ because it is the value that minimises the difference between the distribution of travel time of real trajectories and simulated ones (see Appendix \ref{sec:traf-cal}).
Figure \ref{fig:trajectories} shows four examples of paths generated by OSM, TT, DR with $w=1$ and DR with $w=5$ between the same origin-destination pair (trip).

\begin{figure}[htb!]
    \centering
    \includegraphics[width=0.9\linewidth]{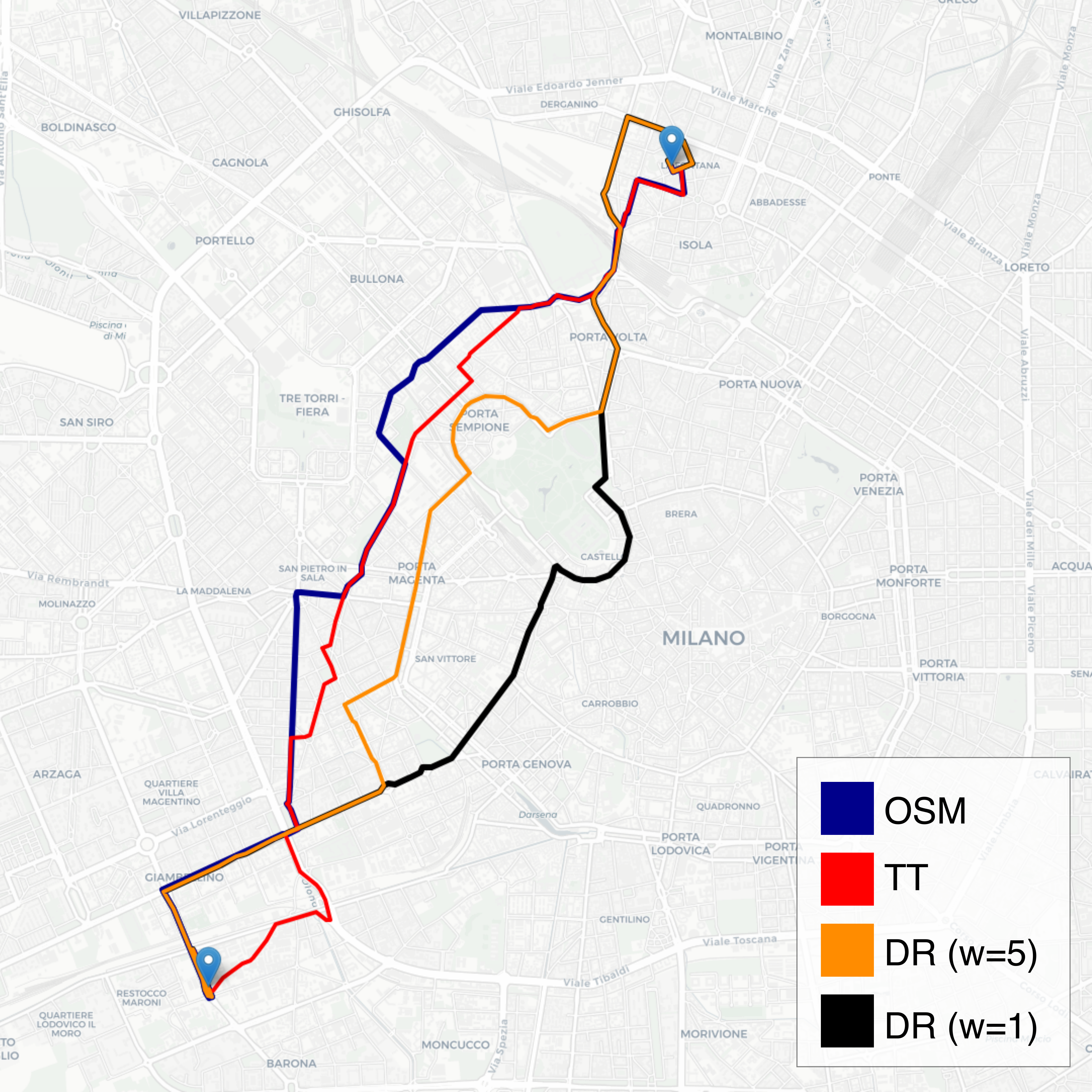}
    \caption{Examples of routed paths between an origin and destination pair according to OSM (blue), TT (red), DR with $w=5$ (orange), and DR with $w=1$ (black).}
    \label{fig:trajectories}
\end{figure}

Given a routing algorithm $R \in \{\text{OSM}, \text{TT}\}$, we study its impact on the urban environment generating 11 multisets of routed paths $\overline{D}^{(R)}_0, \dots, \overline{D}^{(R)}_{10}$. 
In each multiset $\overline{D}^{(R)}_i$ ($i=0, \dots, 10$), $(i \cdot 10)$\% of the paths (chosen uniformly at random among the $N$ paths) are $R$-routed ($R {\in} \{\text{OSM}, \text{TT}\}$) and the remaining paths are routed by DR with $w=5$. 
For example, for $i=5$ and $R = \text{TT}$, $\overline{D}_5^{(R)}$ contains 50\% of the paths routed by TT and the remaining vehicles routed by DR with $w=5$.
Similarly, $i=7$ means that 70\% of the vehicles are TT-routed and 30\% are DR-routed ($w=5$).

To make experiments more robust, for each $i = 0, \dots, 10$, we generate $\overline{D}_i^{(R)}$ ten times, each one with a different choice of $R$-routed vehicles that are chosen uniformly at random.
Finally, we generate the urban traffic $S(\overline{D}_{i}^{(R)}, \mathcal{T})$ for each multiset of routed paths $\overline{D}_{i}^{(R)}$, where departure times in $\mathcal{T}$ are chosen uniformly at random between 0 and 3600 seconds, i.e., we simulate one hour of traffic in Milan. 
Then, for each urban traffic $S(\overline{D}_{i}^{(R)}, \mathcal{T})$, we compute the CO$_2$ emissions for each trajectory and aggregate them at the edge level, obtaining the weighted road network $\overline{G}_i^{(R)}$.
From $\overline{G}_i^{(R)}$, we obtain the overall CO$_2$ emissions in Milan $\mathcal{E}_i^{(R)} = \sum_{e\in E} \mathcal{E}(e)$, where $\mathcal{E}(e)$ is the amount of emissions on edge $e$.

\begin{figure}[htb!]
    \centering
    \subfigure[\large OSM]{
    \includegraphics[width=0.9\linewidth]{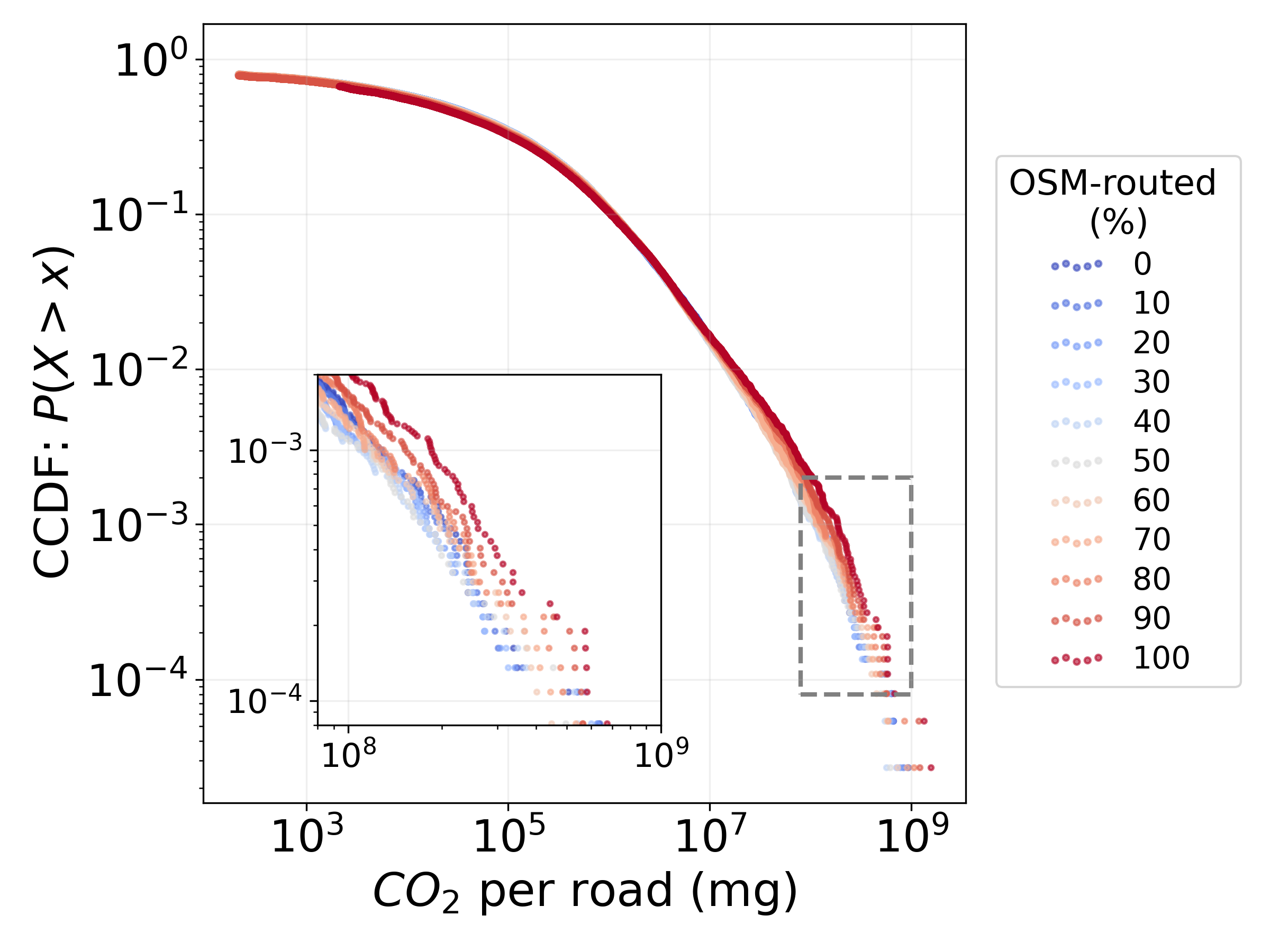}} 
    \subfigure[\large TT]{\includegraphics[width=0.9\linewidth]{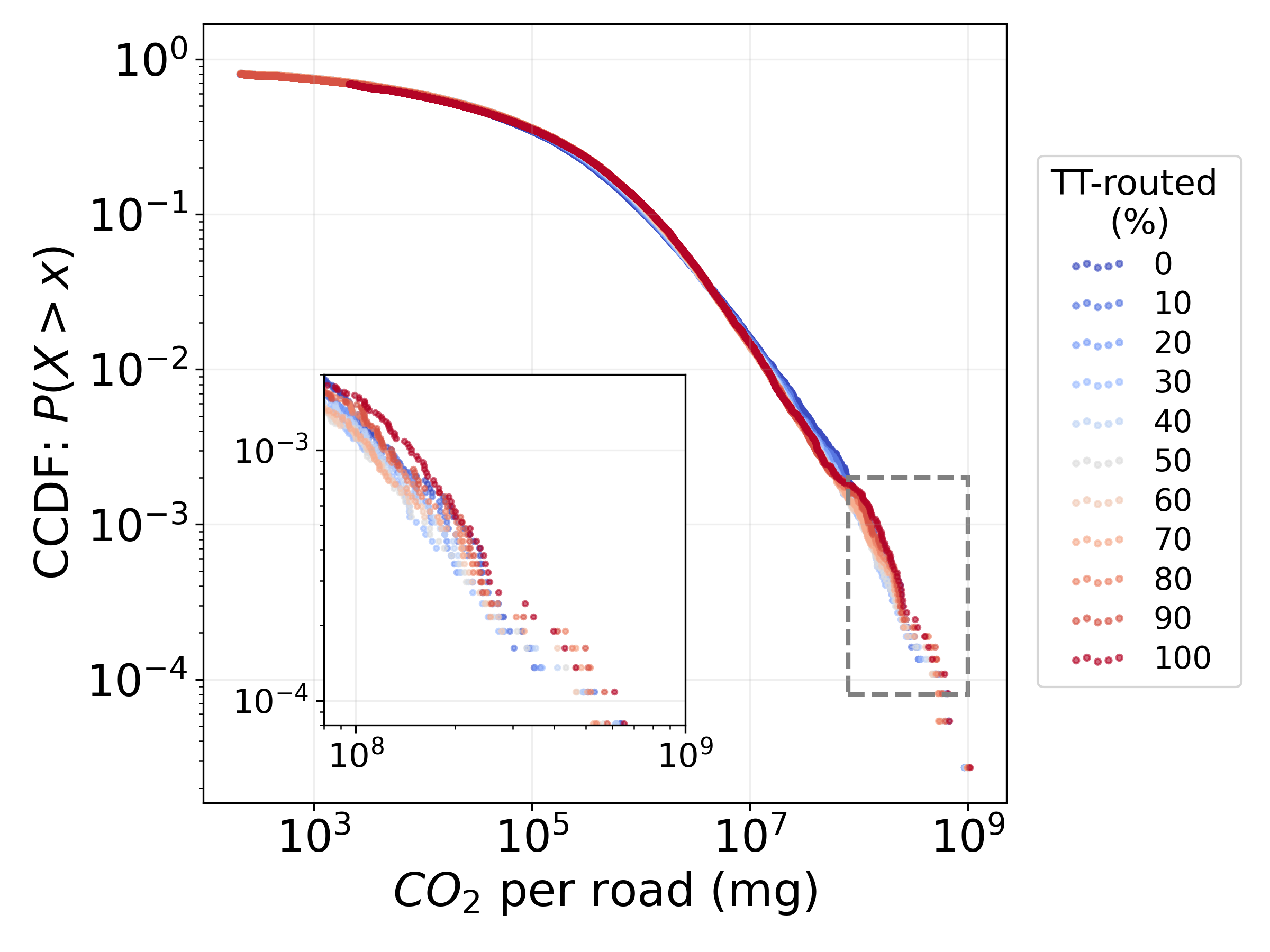}} 
    \caption{
    The Complementary Cumulative Distribution Functions (CCDFs) of the CO$_2$ (in mg) emitted on the roads (averaged across 10 repetitions) by the vehicles of the generated urban traffic $S(\overline{D}_{i}^{(R)}, \mathcal{T})$, for OSM (a) and TT (b). 
    Colours represent routed paths with increasing percentage of $R$-routed vehicles. 
    The inset plot zooms on the distributions' tail.}
    \label{fig:ccdfs}
\end{figure}

\section{Results}
\label{sec:results}
\begin{figure*}[htb!]
    \centering
    \subfigure[\large Gini index]{
    \includegraphics[width=0.8\columnwidth]{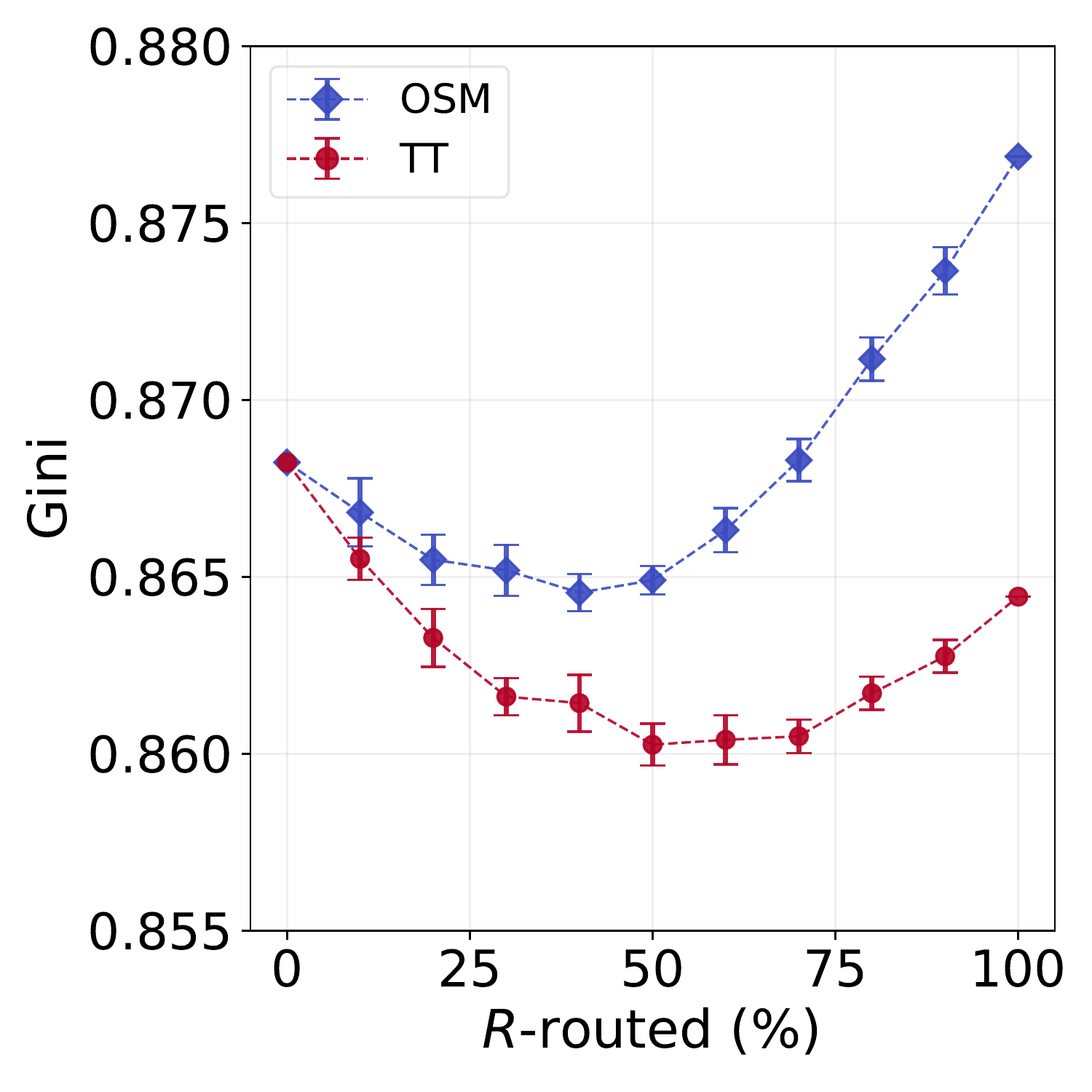}} 
    \subfigure[\large Total CO$_2$]{\includegraphics[width=0.8\columnwidth]{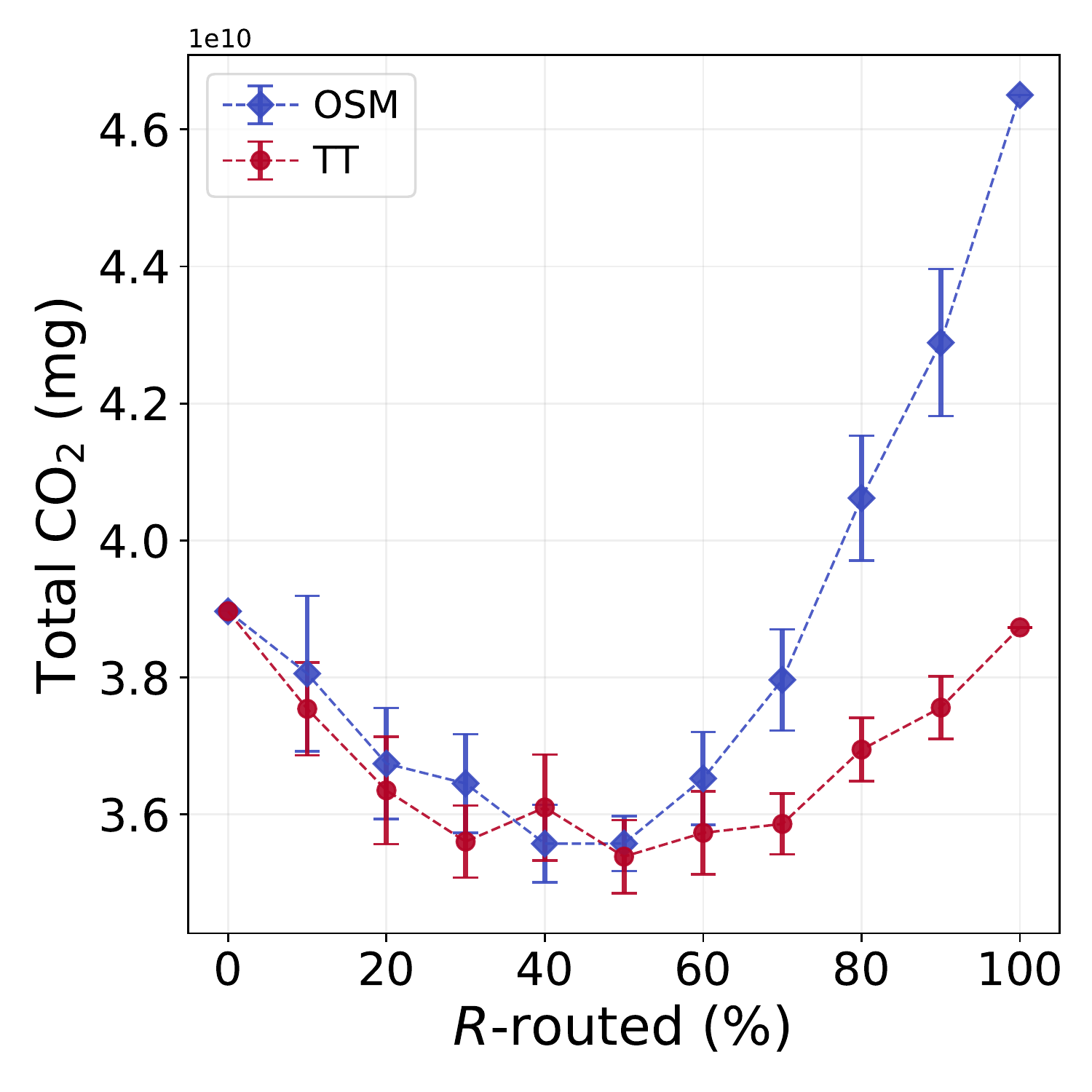}} 
    \caption{
    Gini index of the CO$_2$ distribution (a) and total CO$_2$ emissions (b) varying the percentage of $R$-routed vehicles, for OSM (blue) and TT (red).
    In the error bars, points indicate the average Gini index (a) and the total CO$_2$ (b) over ten simulations with different choices of $R$-routed vehicles chosen uniformly at random.
    Vertical bars indicate the standard deviation. 
    }
    \label{fig:gini&totco2}
\end{figure*}

We study how the distribution of the CO$_2$ emissions across Milan's roads changes varying the percentage of $R$-routed vehicles, i.e., varying $i=0, \dots,10$ of $\overline{D}^{(R)}_i$, $R \in \{\text{OSM}, \text{TT}\}$.
In general, emissions distribute across roads in a heterogeneous way: a few grossly polluted roads coexist with roads with significantly fewer emissions (Figure \ref{fig:ccdfs}).
Indeed, these distributions are associated with a Gini index $g_{\text{\tiny OSM}} \in [0.864, 0.876]$ and $g_{\text{\tiny TT}} \in [0.860, 0.868]$ (Figure~\ref{fig:gini&totco2}a) and are well approximated by a truncated power-law with the exponent $\alpha_{\text{\tiny OSM}} \in [1.76,1.89]$ and $\alpha_{\text{\tiny TT}} \in [1.76,2.00]$ (Figure~\ref{fig:alpha}). 
See Appendix \ref{sec:fitting_distributions} for details on the curve fitting.
In particular, the distributions are the least uneven when 50\% and 70\% of the vehicles are OSM-routed and TT-routed, respectively (Figure~\ref{fig:ccdfs}).

The analysis of how the total CO$_2$ emissions ($\mathcal{E}$) varies with the percentage of $R$-routed vehicles also reveals a clear pattern: when either all vehicles are $R$-routed or none of them, the overall emissions are maximised (Figure \ref{fig:gini&totco2}b).
In contrast, the overall emissions are minimised when just half of the vehicles are R-routed.
In other words, $\mathcal{E}_5^{(R)} < \mathcal{E}_0^{(R)}$ and $\mathcal{E}_5^{(R)} < \mathcal{E}_{10}^{(R)}$.
Note that the total emissions when all vehicles are TT-routed are much heavier than when all are OSM-routed (Figure \ref{fig:gini&totco2}b).
This result suggests that TT's routing algorithm recommends paths that generate a smaller adverse impact on urban well-being than OSM.

\begin{figure*}[htb!]
    \centering
    \subfigure[]{
    \includegraphics[width=0.85\columnwidth]{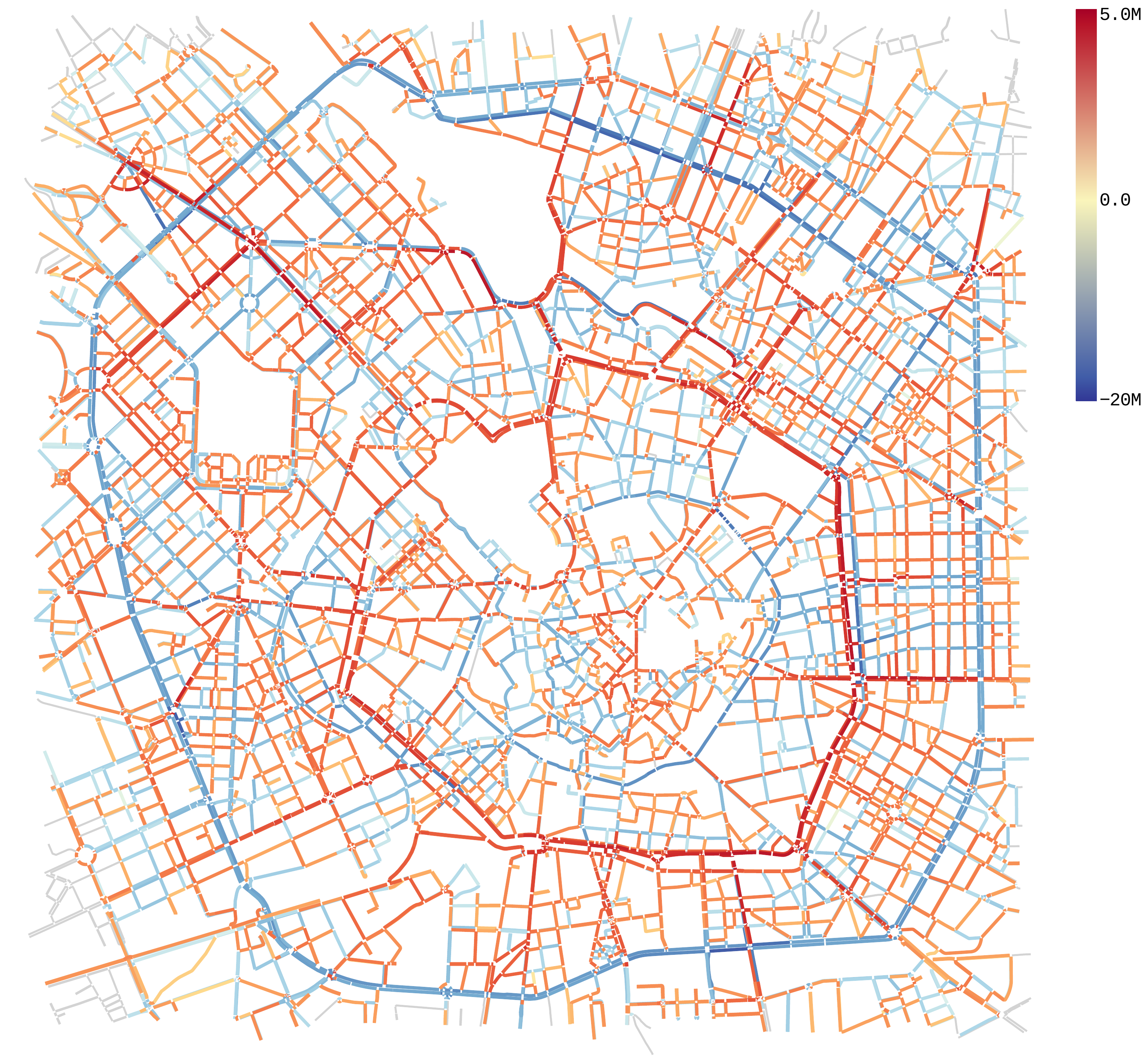}}
    \subfigure[]{\includegraphics[width=0.85\columnwidth]{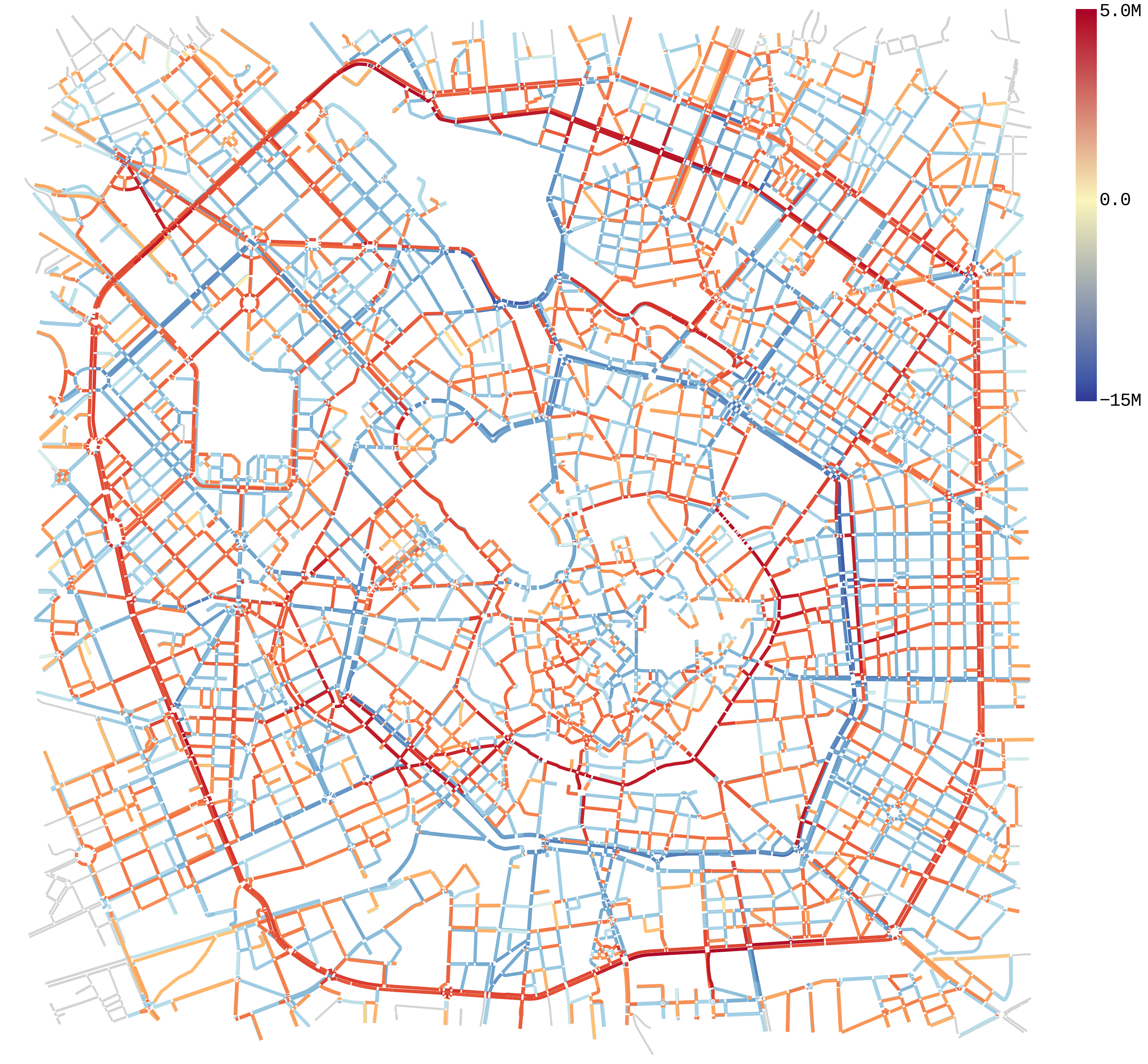}}
    \caption{
    The difference in the total CO$_2$ emitted on each road (in mg per meter of road) when: (a) none of the vehicles is OSM-routed and 50\% of them are
    ($\mathcal{E}_0^{\text{\tiny (OSM)}}(e) - \mathcal{E}_5^{\text{\tiny (OSM)}}(e)$, $\forall e \in E$); (b) all vehicles are OSM-routed and 50\% of them are ($\mathcal{E}_{10}^{\text{\tiny (OSM)}}(e) - \mathcal{E}_5^{\text{\tiny (OSM)}}(e)$, $\forall e \in E$).
    Red roads indicate a positive difference; blue ones indicate a negative difference.
    }
    \label{fig:map_diff}
\end{figure*}

The fraction of $R$-routed vehicles also influence the spatial distribution of emissions in the city.
In particular, comparing the two scenarios that maximise the total emissions (0\% and 100\% of $R$-routed vehicles) with the scenario that minimises them (50\% of $R$-routed vehicles) helps understand where vehicles are being routed, consequently revealing emissions hot spots.
In Figure \ref{fig:map_diff}a, we show the difference between the per-road emissions (normalized by the road length) when none of the vehicle is OSM-routed and 50\% of them are (i.e., $\mathcal{E}_0^{\text{\tiny (OSM)}}(e) - \mathcal{E}_5^{\text{\tiny (OSM)}}(e)$, $\forall e \in E$).
Similarly, Figure \ref{fig:map_diff}b shows the normalized emissions difference when all vehicles are OSM-routed and 50\% of them are (i.e., $\mathcal{E}_{10}^{\text{\tiny(OSM)}}(e) - \mathcal{E}_5^{\text{\tiny (OSM)}}(e)$, $\forall e \in E$).
We find that when 100\% of vehicles are OSM-routed, the emissions are more concentrated towards Milan's ring road (Figure \ref{fig:map_diff}b).
In contrast, when none of them is OSM-routed, the emissions are more concentrated towards the city centre (Figure \ref{fig:map_diff}a).
We find similar results for TT (Figure \ref{fig:map_diff_TT}).

\begin{figure*}[htb!]
    \centering
    \subfigure[\large OSM - Gini]{
	\includegraphics[width=0.3\textwidth]{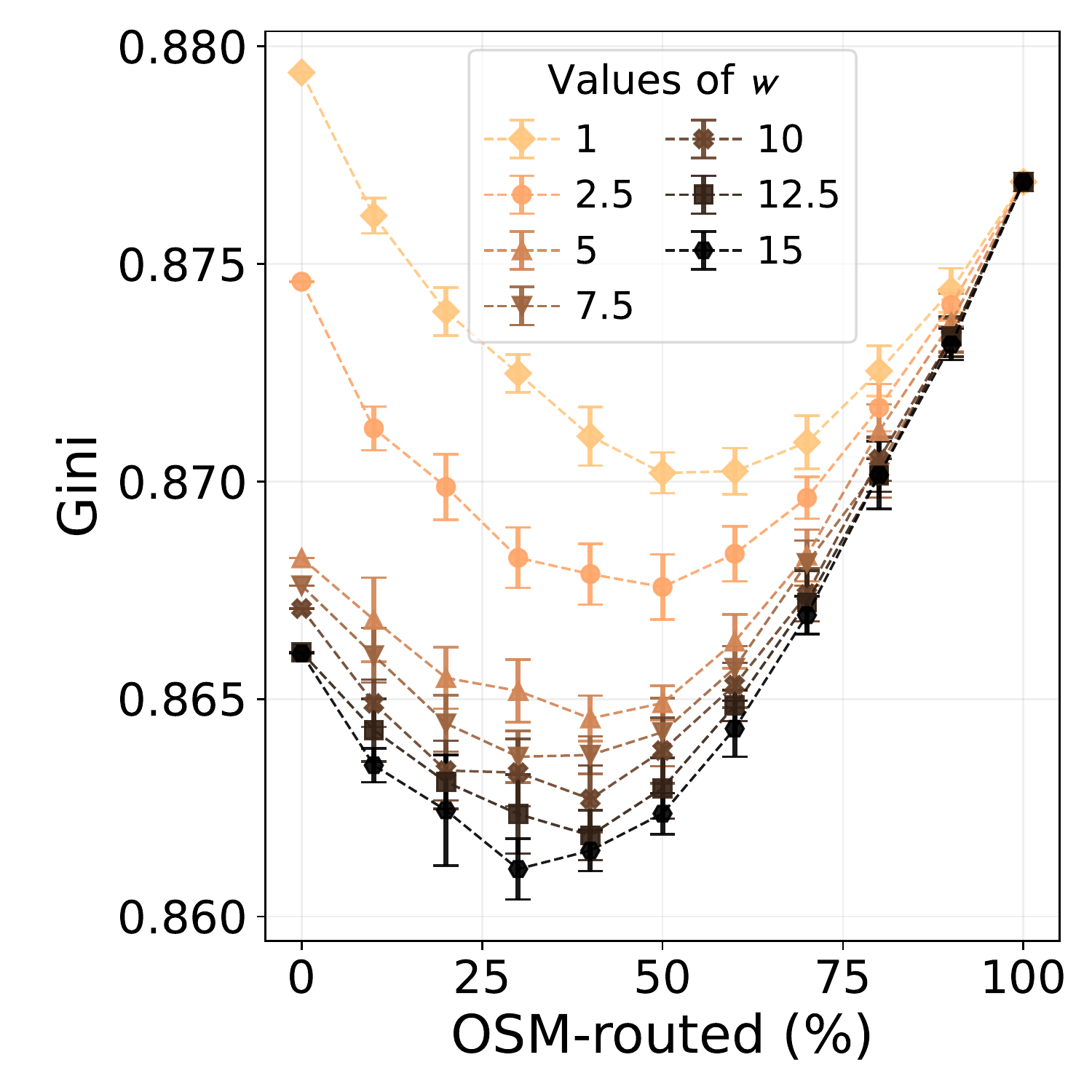}}
	\subfigure[\large OSM - CO$_2$]{
	\includegraphics[width=0.3\textwidth]{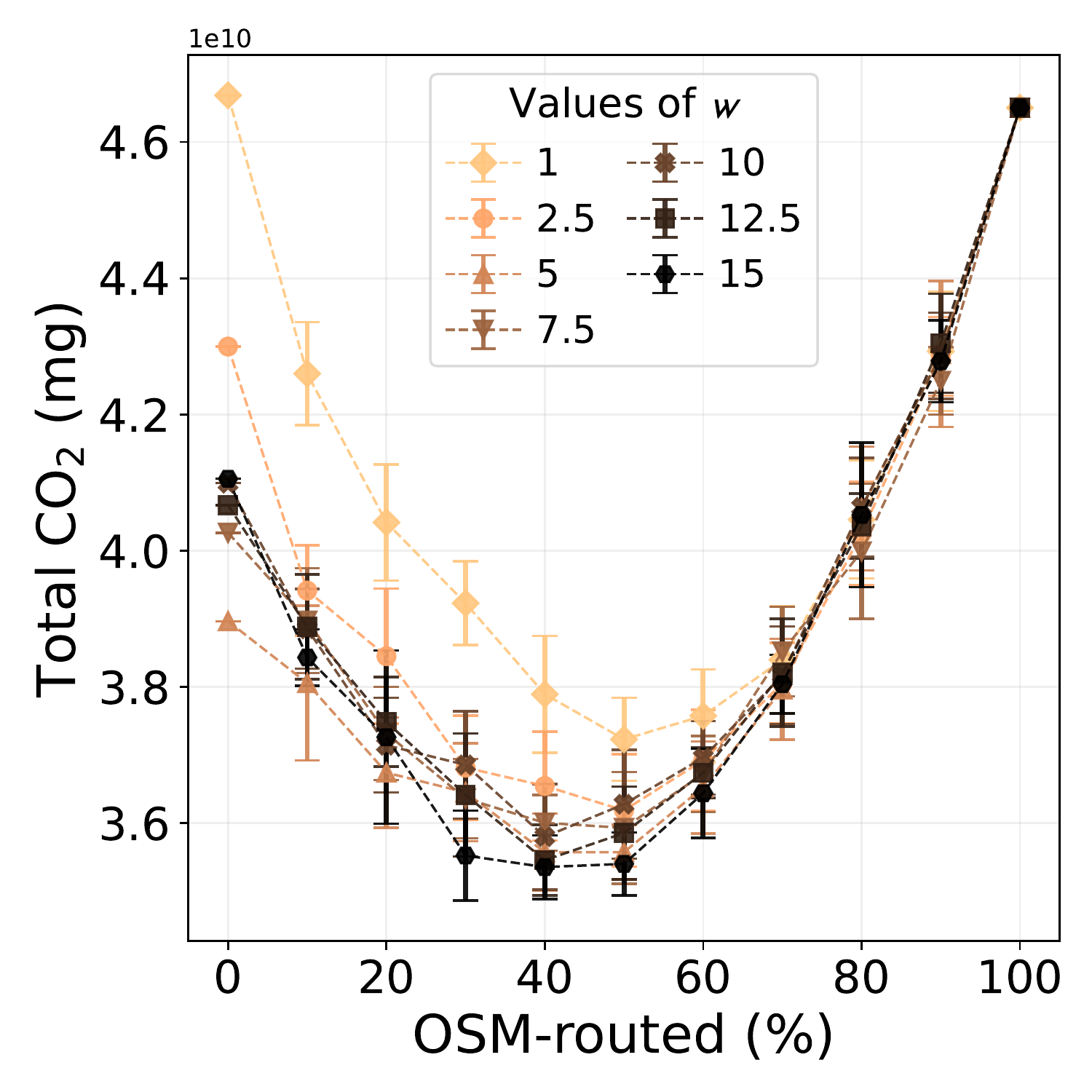}}
	\subfigure[\large OSM - travel time]{
	\includegraphics[width=0.3\textwidth]{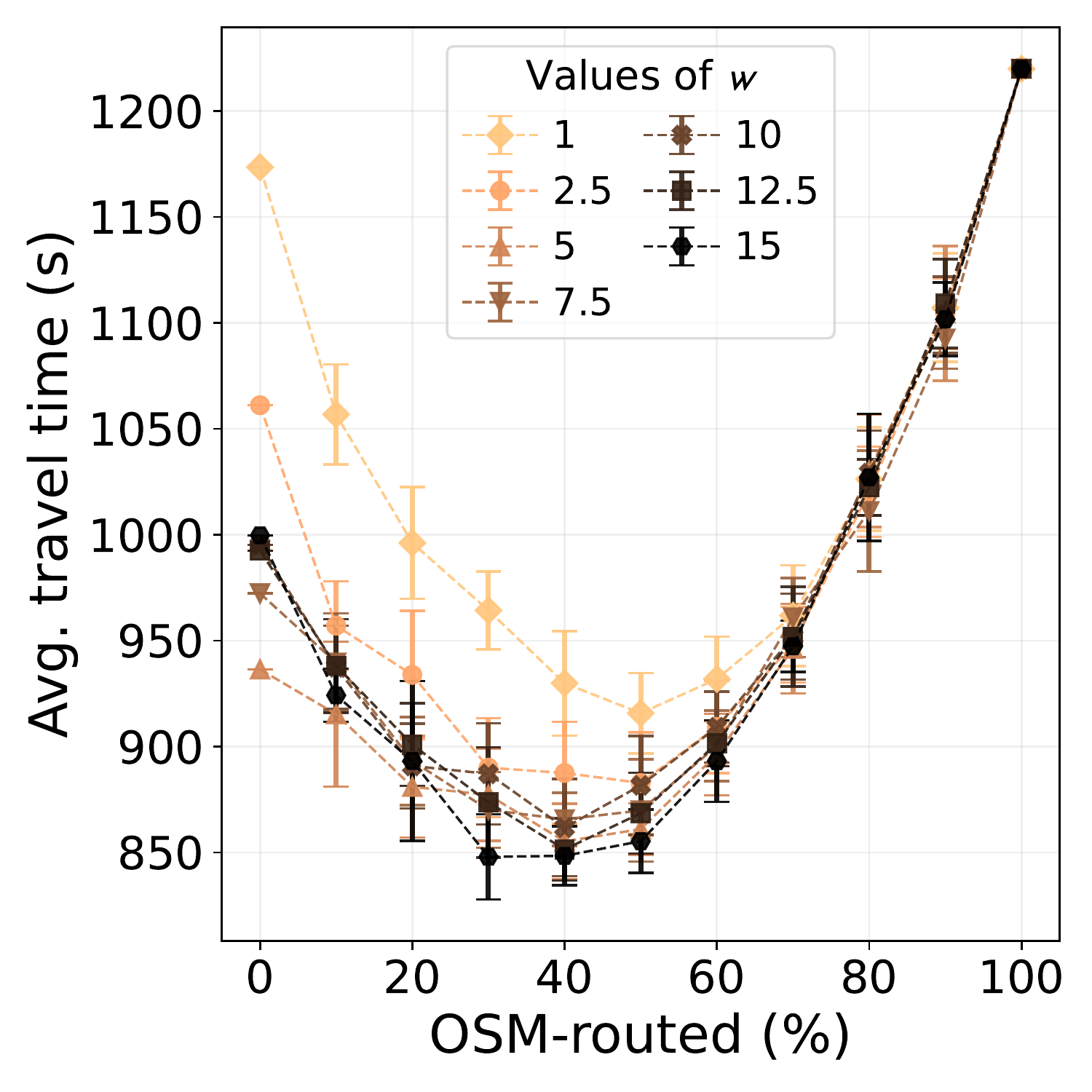}}
	\subfigure[\large TT - Gini]{
	\includegraphics[width=0.3\textwidth]{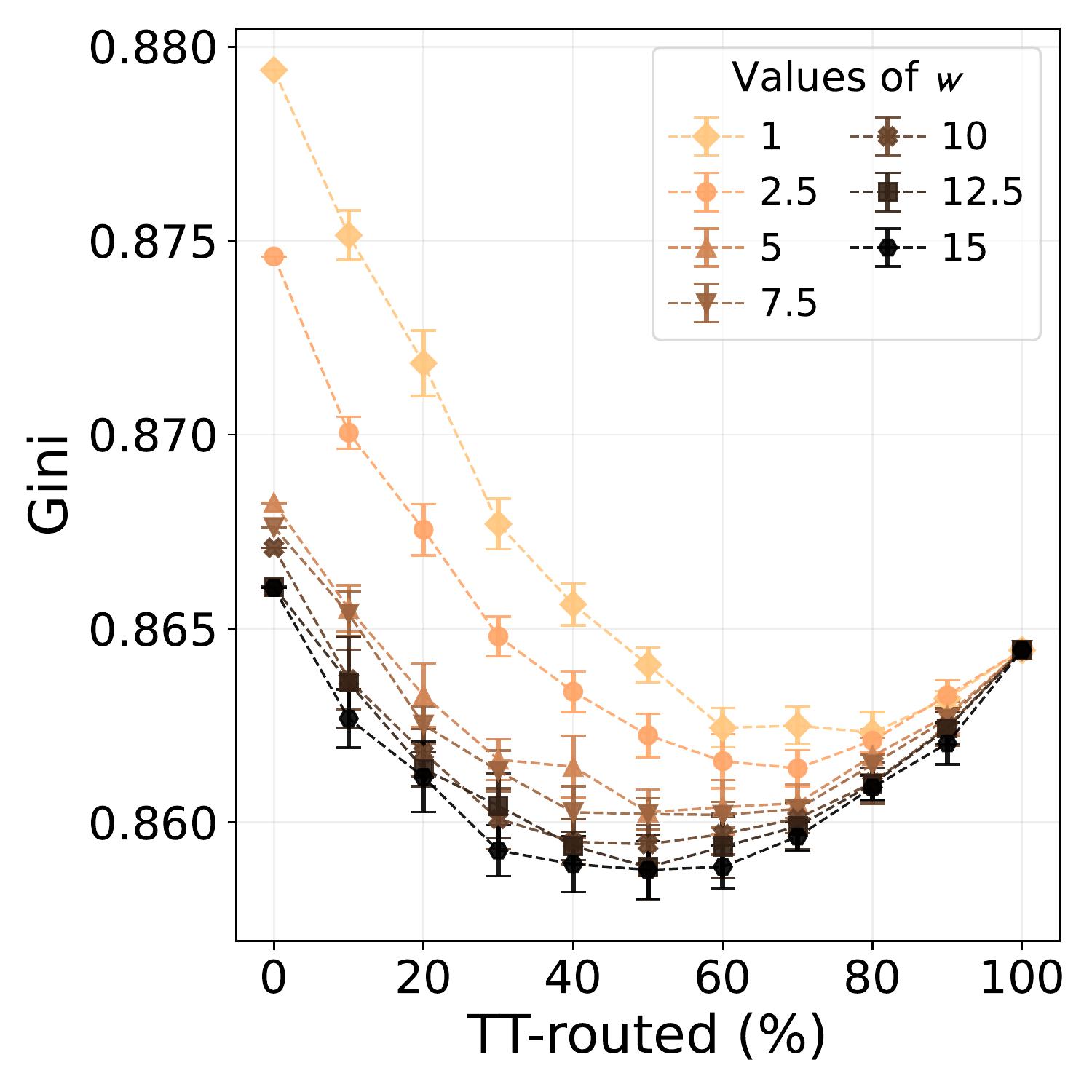}}
	\subfigure[\large TT - CO$_2$]{
	\includegraphics[width=0.3\textwidth]{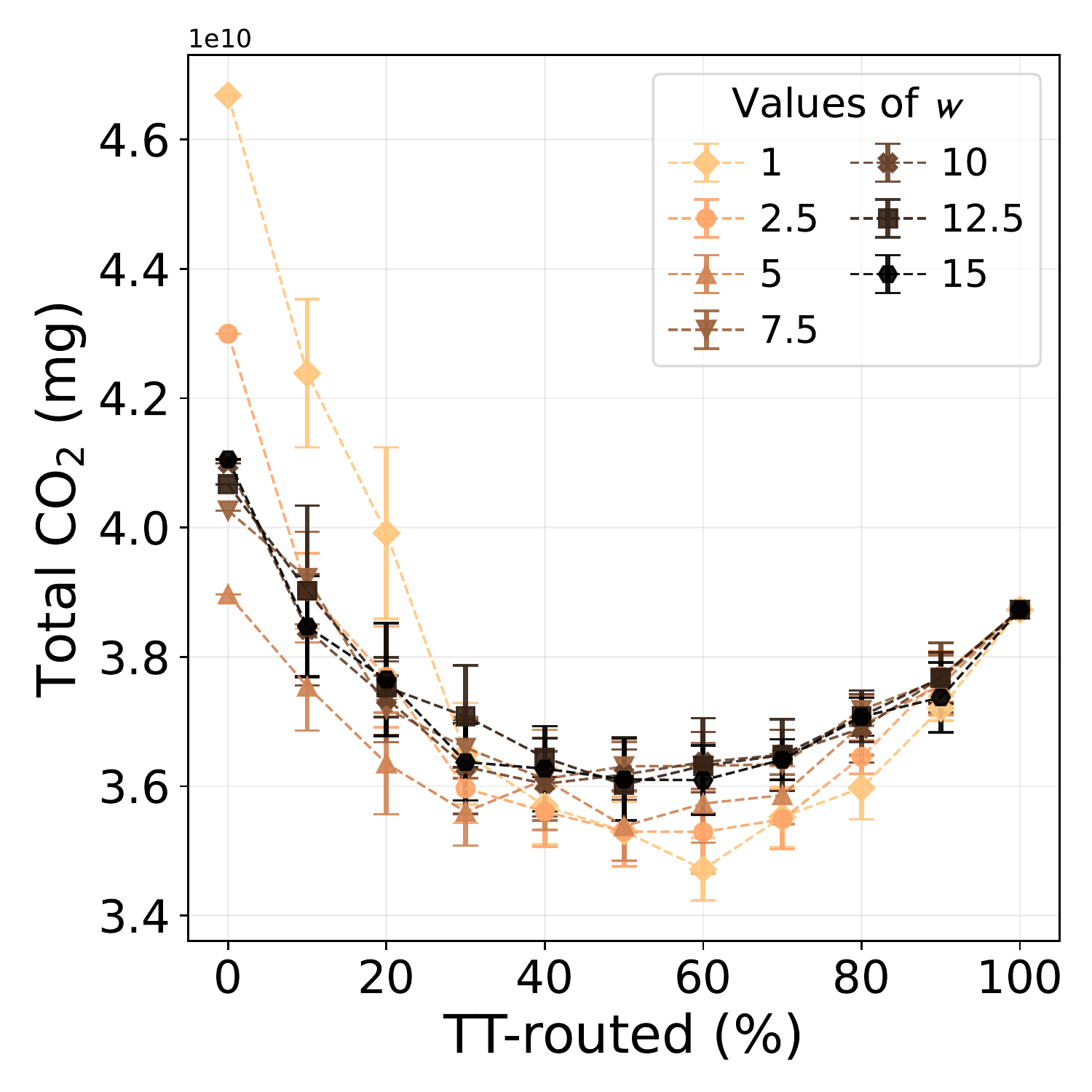}}
	\subfigure[\large TT - travel time]{
	\includegraphics[width=0.3\textwidth]{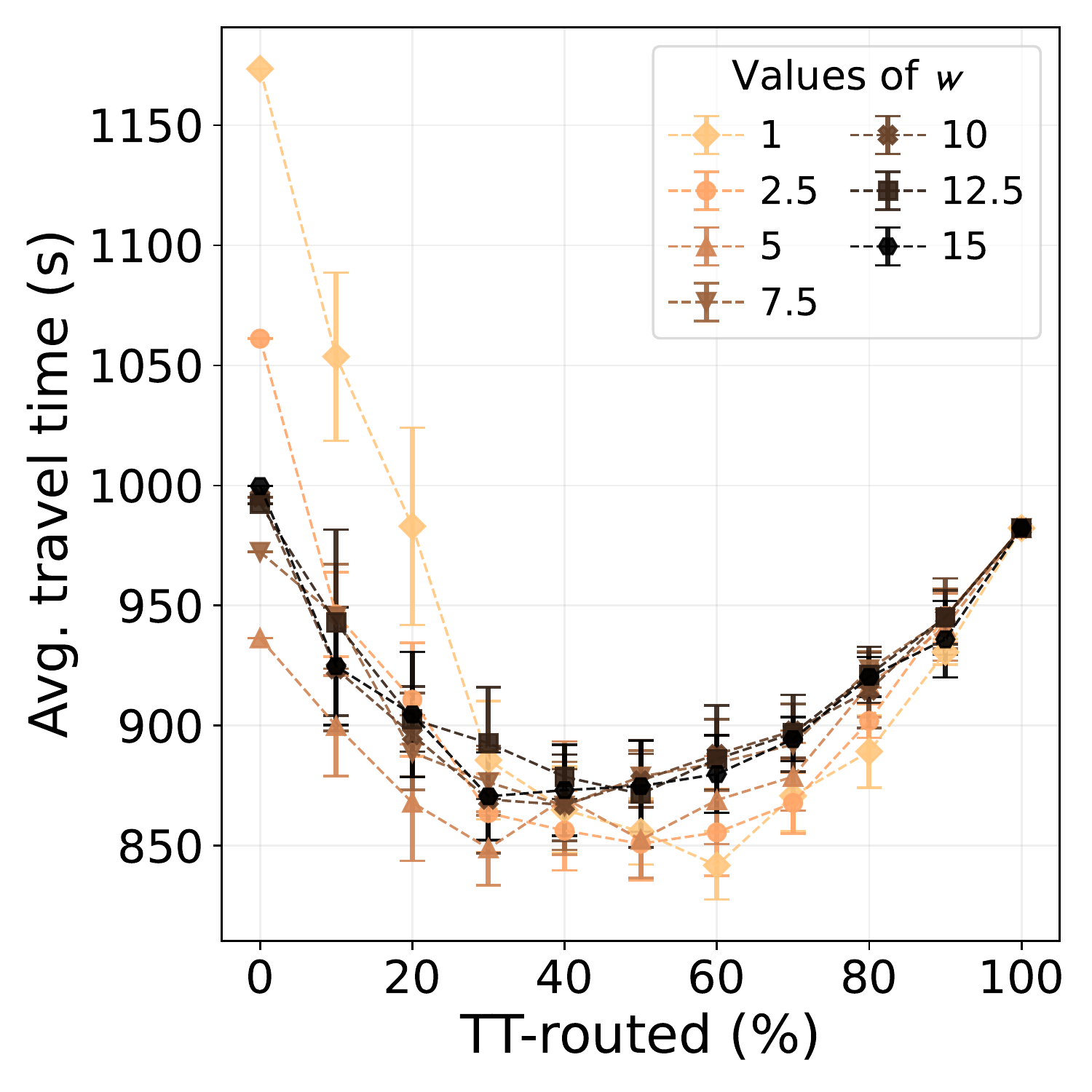}}
    \caption{
    Gini index of CO$_2$ distribution (a,d), total CO$_2$ emissions (b,e), average travel time per vehicle (c,f), for $w = 1, \dots, 15$, varying the percentage of $R$-routed vehicles, for OSM (a,b,c) and TT (d,e,f).
    In the error bars, points indicate the average Gini index (a,d), total CO$_2$ emissions (b,e), average travel time (c,f) over ten simulations with different choices of $R$-routed vehicles chosen uniformly at random.
    Vertical bars indicate the standard deviation.}
    \label{fig:w_osm_tt}
\end{figure*}


\paragraph{Impact of randomization}
We investigate the impact of DR's randomisation parameter $w$ on the total emissions, their distribution over the road network, and the vehicles' average travel time.
For this purpose, we repeat the simulation of the urban traffic varying $w = 1, \dots, 15$. 
We remind that the higher $w$, the more randomly perturbed DR's fastest path is (see Appendix \ref{sec:perturbation}).

First, changing $w$ does not affect considerably the shape of the emission distributions: the Gini index still ranges in $g_{\text{\tiny OSM}} \in [0.860, 0.879]$ and $g_{\text{\tiny TT}} \in [0.858, 0.879]$ (Figure \ref{fig:w_osm_tt}a, d) and the truncated power-law exponent in $\alpha_{\text{\tiny OSM}} \in [1.71,1.94]$ and $\alpha_{\text{\tiny TT}} \in [1.71,2.08]$.
Second, the higher $w$, the more even the distributions (Figure \ref{fig:w_osm_tt}a, d), the lower the emissions (Figure \ref{fig:w_osm_tt}b, e), and the shorter the average travel time (Figure \ref{fig:w_osm_tt}c, f).
Third, the total emissions are minimised when 40-60\% of the vehicles are $R$-routed, and so are the average travel time and the distributions' Gini index (Figure \ref{fig:w_osm_tt}).

\section{Discussion}
Our study provide several interesting results. 
We discuss and interpret them in the following paragraphs.

\textbf{Navigation apps do impact on emissions.} 
In general, CO$_2$ emissions are unevenly distributed across the city roads, and this unevenness is exacerbated when all vehicles or none of them are $R$-routed. 
In contrast, the total CO$_2$ emissions as well as their heterogeneity are minimised when around 50-70\% of the vehicles are $R$-routed. 
These results clearly show that navigation apps do have a non-negligible adverse impact on the urban environment.

\textbf{Path perturbation is beneficial.}
We also find that path perturbation (through DR's $w$ parameter) is beneficial: the more we randomise the vehicles' paths, the shorter their travel time and the lower the amount of emissions in the city.
This may be because the vehicles' perturbed paths are more ``diverse'' (i.e., they spread over more roads), thus reducing congestion and consequently the amount of emissions and travel time.
The role of path randomisation is interesting and deserves further investigation.


\textbf{Trips' spatial distribution matters.} 
Urban planners and policymakers may be interested in investigating the impact of navigation apps on the spatial distribution of vehicles' emissions. 
Our study shows that, in Milan, the more vehicles are $R$-routed, the less emissions concentrate in the city center and the more in the external ring road.
Presumably, this may be because the navigation apps 
route the vehicles preferably on the city's arterial roads (such as the ring road) to keep the individual path as fast as possible.

\textbf{Different navigation apps, different impact.} 
Our TraffiCO$_2$ simulation framework is a useful tool to compare the impact of different navigation apps on urban well-being. In our study, we find that TomTom is better than OpenStreetMap in minimising the emissions distribution's inequality, and the city's total emissions and average travel time. 
Although TomTom's algorithm is not public, we may suppose that TomTom's heuristics are more sophisticated, e.g., they use more or higher-quality information than OpenStreetMap, such as traffic data, detailed information about typical road speed, capacity, and length.

\textbf{A novel what-if analysis tool.} 
TraffiCO$_2$ is a simulation framework to compare the impact of different routing strategies on urban well-being in terms of amount of emissions and their spatial and statistical distribution. 
Our tool allows us to go beyond state-of-the-art studies, which investigate the reduction of emissions for a fixed fraction of vehicles routed by a navigation app ~\cite{arora2021quantifying}.

\section{Conclusion}
\label{sec:conclusion}
Scenarios where either all vehicles or none of them follow a navigation app's suggestion lead to the highest CO$_2$ emissions and the most uneven distribution of emissions per road.
We show that when just a fraction of the vehicles (around half of them) follows the navigation app's suggestions, such an adverse impact is minimised.
This minimisation also holds when introducing more randomness in non-R-routed paths, which leads to a reduction of the vehicles' average travel time, overall CO$_2$ emissions, and inequality in the distribution of CO$_2$ emissions across the road network.

We plan to improve and extend this study in several directions. 
First and foremost, we plan to extend the set of navigation apps considered (e.g., to Google Maps) and to study the environmental impact of a fleet of vehicles that use various navigation apps.
In this work, we focus on light-duty vehicles only and assume that all vehicles carry the same engine type. 
Including heavy-duty vehicles (e.g., buses, trucks) and considering different vehicle engine ages and types (e.g. diesel, LPG, petrol) would provide a more complete mosaic of the emissions on the road network.
We also look forward to apply our framework to various cities, to investigate how the impact of navigation apps varies with city size, shape, road network, and other characteristics.
Moreover, we could study the impact of navigation apps in terms of other pollutants, such as nitrogen oxides, ozone, particulate matter, and volatile organic compounds, using emissions models similar to those used in this paper.

In the meantime, our work is a first step towards designing next-generation routing algorithms that, as our results suggest, should consider some degree of path randomisation to increase urban well-being while still satisfying individual needs.


\begin{acks}
This work has been partially supported by EU project H2020 Humane-AI-net G.A. 952026, EU project H2020 SoBigData++ G.A. 871042, ERC-2018-ADG G.A. 834756 “XAI: Science and technology for the eXplanation of AI decision making”, and by the CHIST-ERA grant CHIST-ERA-19-XAI-010, by MUR (grant No. not yet available), FWF (grant No. I 5205), EPSRC (grant No. EP/V055712/1), NCN (grant No. 2020/02/Y/ST6/00064), ETAg (grant No. SLTAT21096), BNSF (grant No. K$\Pi$-06-AOO2/5). 

We thank Davide Nicola, Pietro Iemmello, René Ferretti, Dante Milonga and Maurizio Turone for the inspiration. 
We thank Daniele Fadda for his precious
support in data visualization.
\end{acks}

\bibliographystyle{ACM-Reference-Format}
\bibliography{biblio}


\begin{thebibliography}{63}


\ifx \showCODEN    \undefined \def \showCODEN     #1{\unskip}     \fi
\ifx \showDOI      \undefined \def \showDOI       #1{#1}\fi
\ifx \showISBNx    \undefined \def \showISBNx     #1{\unskip}     \fi
\ifx \showISBNxiii \undefined \def \showISBNxiii  #1{\unskip}     \fi
\ifx \showISSN     \undefined \def \showISSN      #1{\unskip}     \fi
\ifx \showLCCN     \undefined \def \showLCCN      #1{\unskip}     \fi
\ifx \shownote     \undefined \def \shownote      #1{#1}          \fi
\ifx \showarticletitle \undefined \def \showarticletitle #1{#1}   \fi
\ifx \showURL      \undefined \def \showURL       {\relax}        \fi
\providecommand\bibfield[2]{#2}
\providecommand\bibinfo[2]{#2}
\providecommand\natexlab[1]{#1}
\providecommand\showeprint[2][]{arXiv:#2}

\bibitem[Ahn and Rakha(2013)]%
        {ahn2013network}
\bibfield{author}{\bibinfo{person}{Kyoungho Ahn} {and}
  \bibinfo{person}{Hesham~A. Rakha}.} \bibinfo{year}{2013}\natexlab{}.
\newblock \showarticletitle{Network-wide impacts of eco-routing strategies: A
  large-scale case study}.
\newblock \bibinfo{journal}{\emph{Transportation Research Part D: Transport and
  Environment}}  \bibinfo{volume}{25} (\bibinfo{year}{2013}),
  \bibinfo{pages}{119--130}.
\newblock
\showISSN{1361-9209}
\urldef\tempurl%
\url{https://doi.org/10.1016/j.trd.2013.09.006}
\showDOI{\tempurl}


\bibitem[Alazzawi et~al\mbox{.}(2018)]%
        {alazzawi2018simulating}
\bibfield{author}{\bibinfo{person}{Sabina Alazzawi}, \bibinfo{person}{Mathias
  Hummel}, \bibinfo{person}{Pascal Kordt}, \bibinfo{person}{Thorsten
  Sickenberger}, \bibinfo{person}{Christian Wieseotte}, {and}
  \bibinfo{person}{Oliver Wohak}.} \bibinfo{year}{2018}\natexlab{}.
\newblock \showarticletitle{Simulating the Impact of Shared, Autonomous
  Vehicles on Urban Mobility -- a Case Study of Milan}. In
  \bibinfo{booktitle}{\emph{SUMO 2018- Simulating Autonomous and Intermodal
  Transport Systems}} \emph{(\bibinfo{series}{EPiC Series in Engineering},
  Vol.~\bibinfo{volume}{2})}, \bibfield{editor}{\bibinfo{person}{Evamarie
  Wie\{\textbackslash{}ss\}ner}, \bibinfo{person}{Leonhard
  L\textbackslash{}"ucken}, \bibinfo{person}{Robert Hilbrich},
  \bibinfo{person}{Yun-Pang Fl\textbackslash{}"otter\textbackslash{}"od},
  \bibinfo{person}{Jakob Erdmann}, \bibinfo{person}{Laura Bieker-Walz}, {and}
  \bibinfo{person}{Michael Behrisch}} (Eds.). \bibinfo{publisher}{EasyChair},
  \bibinfo{pages}{94--110}.
\newblock
\showISSN{2516-2330}
\urldef\tempurl%
\url{https://doi.org/10.29007/2n4h}
\showDOI{\tempurl}


\bibitem[Alstott et~al\mbox{.}(2014)]%
        {alstott2013powerlaw}
\bibfield{author}{\bibinfo{person}{Jeff Alstott}, \bibinfo{person}{Ed
  Bullmore}, {and} \bibinfo{person}{Dietmar Plenz}.}
  \bibinfo{year}{2014}\natexlab{}.
\newblock \showarticletitle{powerlaw: A Python Package for Analysis of
  Heavy-Tailed Distributions}.
\newblock \bibinfo{journal}{\emph{PLOS ONE}} \bibinfo{volume}{9},
  \bibinfo{number}{1} (\bibinfo{date}{01} \bibinfo{year}{2014}),
  \bibinfo{pages}{1--11}.
\newblock
\urldef\tempurl%
\url{https://doi.org/10.1371/journal.pone.0085777}
\showDOI{\tempurl}


\bibitem[Argota Sánchez-Vaquerizo(2022)]%
        {Argota2022Getting}
\bibfield{author}{\bibinfo{person}{Javier Argota Sánchez-Vaquerizo}.}
  \bibinfo{year}{2022}\natexlab{}.
\newblock \showarticletitle{Getting Real: The Challenge of Building and
  Validating a Large-Scale Digital Twin of Barcelona's Traffic with Empirical
  Data}.
\newblock \bibinfo{journal}{\emph{ISPRS International Journal of
  Geo-Information}} \bibinfo{volume}{11}, \bibinfo{number}{1}
  (\bibinfo{year}{2022}).
\newblock
\showISSN{2220-9964}
\urldef\tempurl%
\url{https://doi.org/10.3390/ijgi11010024}
\showDOI{\tempurl}


\bibitem[Arora et~al\mbox{.}(2021)]%
        {arora2021quantifying}
\bibfield{author}{\bibinfo{person}{Neha Arora}, \bibinfo{person}{Theophile
  Cabannes}, \bibinfo{person}{Sanjay~Ganapathy Subramaniam},
  \bibinfo{person}{Yechen Li}, \bibinfo{person}{Preston McAfee},
  \bibinfo{person}{Marc Nunkesser}, \bibinfo{person}{Carolina Osorio},
  \bibinfo{person}{Andrew Tomkins}, {and} \bibinfo{person}{Iveel Tsogsuren}.}
  \bibinfo{year}{2021}\natexlab{}.
\newblock \showarticletitle{Quantifying the sustainability impact of Google
  Maps: A case study of Salt Lake City}.
\newblock


\bibitem[Aziz and Ukkusuri(2018)]%
        {aziz2018novel}
\bibfield{author}{\bibinfo{person}{H.~M.~Abdul Aziz} {and}
  \bibinfo{person}{Satish~V. Ukkusuri}.} \bibinfo{year}{2018}\natexlab{}.
\newblock \showarticletitle{A novel approach to estimate emissions from large
  transportation networks: Hierarchical clustering-based link-driving-schedules
  for EPA-MOVES using dynamic time warping measures}.
\newblock \bibinfo{journal}{\emph{International Journal of Sustainable
  Transportation}} \bibinfo{volume}{12}, \bibinfo{number}{3}
  (\bibinfo{year}{2018}), \bibinfo{pages}{192--204}.
\newblock
\urldef\tempurl%
\url{https://doi.org/10.1080/15568318.2017.1346732}
\showDOI{\tempurl}


\bibitem[Bai et~al\mbox{.}(2018)]%
        {bai2018six}
\bibfield{author}{\bibinfo{person}{Xuemei Bai}, \bibinfo{person}{Richard
  Dawson}, \bibinfo{person}{Diana Ürge Vorsatz}, \bibinfo{person}{Gian
  Delgado}, \bibinfo{person}{Aliyu~Salisu Barau}, \bibinfo{person}{Shobhakar
  Dhakal}, \bibinfo{person}{David Dodman}, \bibinfo{person}{Lykke Leonardsen},
  \bibinfo{person}{Valerie Masson-Delmotte}, \bibinfo{person}{Debra Roberts},
  {and} \bibinfo{person}{Seth Schultz}.} \bibinfo{year}{2018}\natexlab{}.
\newblock \showarticletitle{Six Research Priorities for Cities and Climate
  Change}.
\newblock \bibinfo{journal}{\emph{Nature}}  \bibinfo{volume}{555}
  (\bibinfo{date}{02} \bibinfo{year}{2018}).
\newblock
\urldef\tempurl%
\url{https://doi.org/10.1038/d41586-018-02409-z}
\showDOI{\tempurl}


\bibitem[Barbosa et~al\mbox{.}(2018)]%
        {barbosa2018human}
\bibfield{author}{\bibinfo{person}{Hugo Barbosa}, \bibinfo{person}{Marc
  Barthelemy}, \bibinfo{person}{Gourab Ghoshal}, \bibinfo{person}{Charlotte~R
  James}, \bibinfo{person}{Maxime Lenormand}, \bibinfo{person}{Thomas Louail},
  \bibinfo{person}{Ronaldo Menezes}, \bibinfo{person}{Jos{\'e}~J Ramasco},
  \bibinfo{person}{Filippo Simini}, {and} \bibinfo{person}{Marcello Tomasini}.}
  \bibinfo{year}{2018}\natexlab{}.
\newblock \showarticletitle{Human mobility: Models and applications}.
\newblock \bibinfo{journal}{\emph{Physics Reports}}  \bibinfo{volume}{734}
  (\bibinfo{year}{2018}), \bibinfo{pages}{1--74}.
\newblock


\bibitem[Barth et~al\mbox{.}(2007)]%
        {barth2007environmentally}
\bibfield{author}{\bibinfo{person}{Matthew Barth}, \bibinfo{person}{Kanok
  Boriboonsomsin}, {and} \bibinfo{person}{Alex Vu}.}
  \bibinfo{year}{2007}\natexlab{}.
\newblock \showarticletitle{Environmentally-Friendly Navigation}. In
  \bibinfo{booktitle}{\emph{2007 IEEE Intelligent Transportation Systems
  Conference}}. \bibinfo{pages}{684--689}.
\newblock
\urldef\tempurl%
\url{https://doi.org/10.1109/ITSC.2007.4357672}
\showDOI{\tempurl}


\bibitem[Besse et~al\mbox{.}(2015)]%
        {Review2015Besse}
\bibfield{author}{\bibinfo{person}{Philippe Besse}, \bibinfo{person}{Brendan
  Guillouet}, \bibinfo{person}{Jean-Michel Loubes}, {and}
  \bibinfo{person}{Francois Royer}.} \bibinfo{year}{2015}\natexlab{}.
\newblock \showarticletitle{Review \& Perspective for Distance Based Trajectory
  Clustering}.
\newblock  (\bibinfo{date}{08} \bibinfo{year}{2015}).
\newblock


\bibitem[B{\"{o}}hm et~al\mbox{.}(2022)]%
        {bohm2021improving}
\bibfield{author}{\bibinfo{person}{Matteo B{\"{o}}hm}, \bibinfo{person}{Mirco
  Nanni}, {and} \bibinfo{person}{Luca Pappalardo}.}
  \bibinfo{year}{2022}\natexlab{}.
\newblock \showarticletitle{{Gross polluters and vehicle emissions reduction}}.
\newblock \bibinfo{journal}{\emph{Nature Sustainability}}
  (\bibinfo{year}{2022}).
\newblock
\showISSN{2398-9629}
\urldef\tempurl%
\url{https://doi.org/10.1038/s41893-022-00903-x}
\showDOI{\tempurl}


\bibitem[Bongiorno et~al\mbox{.}(2021)]%
        {bongiorno2021vector}
\bibfield{author}{\bibinfo{person}{Christian Bongiorno}, \bibinfo{person}{Yulun
  Zhou}, \bibinfo{person}{Marta Kryven}, \bibinfo{person}{David Theurel},
  \bibinfo{person}{Alessandro Rizzo}, \bibinfo{person}{Paolo Santi},
  \bibinfo{person}{Joshua Tenenbaum}, {and} \bibinfo{person}{Carlo Ratti}.}
  \bibinfo{year}{2021}\natexlab{}.
\newblock \showarticletitle{Vector-based pedestrian navigation in cities}.
\newblock \bibinfo{journal}{\emph{Nature Computational Science}}
  \bibinfo{volume}{1} (\bibinfo{year}{2021}), \bibinfo{pages}{678–685}.
\newblock
\urldef\tempurl%
\url{https://doi.org/10.1038/s43588-021-00130-y}
\showDOI{\tempurl}


\bibitem[Cervero and Kockelman(1997)]%
        {cervero1997travel}
\bibfield{author}{\bibinfo{person}{Robert Cervero} {and} \bibinfo{person}{Kara
  Kockelman}.} \bibinfo{year}{1997}\natexlab{}.
\newblock \showarticletitle{Travel demand and the 3Ds: Density, diversity, and
  design}.
\newblock \bibinfo{journal}{\emph{Transportation Research Part D: Transport and
  Environment}} \bibinfo{volume}{2}, \bibinfo{number}{3}
  (\bibinfo{year}{1997}), \bibinfo{pages}{199--219}.
\newblock
\showISSN{1361-9209}
\urldef\tempurl%
\url{https://doi.org/10.1016/S1361-9209(97)00009-6}
\showDOI{\tempurl}


\bibitem[Chatterton et~al\mbox{.}(2015)]%
        {chatterton2015use}
\bibfield{author}{\bibinfo{person}{Tim Chatterton}, \bibinfo{person}{Jo
  Barnes}, \bibinfo{person}{R.~Eddie Wilson}, \bibinfo{person}{Jillian Anable},
  {and} \bibinfo{person}{Sally Cairns}.} \bibinfo{year}{2015}\natexlab{}.
\newblock \showarticletitle{Use of a novel dataset to explore spatial and
  social variations in car type, size, usage and emissions}.
\newblock \bibinfo{journal}{\emph{Transportation Research Part D: Transport and
  Environment}}  \bibinfo{volume}{39} (\bibinfo{year}{2015}),
  \bibinfo{pages}{151--164}.
\newblock
\showISSN{1361-9209}
\urldef\tempurl%
\url{https://doi.org/10.1016/j.trd.2015.06.003}
\showDOI{\tempurl}


\bibitem[Chen et~al\mbox{.}(2020)]%
        {chen2020mining}
\bibfield{author}{\bibinfo{person}{Jinyu Chen}, \bibinfo{person}{Wenjing Li},
  \bibinfo{person}{Haoran Zhang}, \bibinfo{person}{Wenxiao Jiang},
  \bibinfo{person}{Weifeng Li}, \bibinfo{person}{Yi Sui}, \bibinfo{person}{Xuan
  Song}, {and} \bibinfo{person}{Ryosuke Shibasaki}.}
  \bibinfo{year}{2020}\natexlab{}.
\newblock \showarticletitle{Mining urban sustainable performance: GPS
  data-based spatio-temporal analysis on on-road braking emission}.
\newblock \bibinfo{journal}{\emph{Journal of Cleaner Production}}
  \bibinfo{volume}{270} (\bibinfo{year}{2020}), \bibinfo{pages}{122489}.
\newblock
\showISSN{0959-6526}
\urldef\tempurl%
\url{https://doi.org/10.1016/j.jclepro.2020.122489}
\showDOI{\tempurl}


\bibitem[Chong et~al\mbox{.}(2020)]%
        {chong2020}
\bibfield{author}{\bibinfo{person}{Hwan~S. Chong}, \bibinfo{person}{Sangil
  Kwon}, \bibinfo{person}{Yunsung Lim}, {and} \bibinfo{person}{Jongtae Lee}.}
  \bibinfo{year}{2020}\natexlab{}.
\newblock \showarticletitle{Real-world fuel consumption, gaseous pollutants,
  and CO2 emission of light-duty diesel vehicles}.
\newblock \bibinfo{journal}{\emph{Sustainable Cities and Society}}
  \bibinfo{volume}{53} (\bibinfo{year}{2020}), \bibinfo{pages}{101925}.
\newblock
\showISSN{2210-6707}
\urldef\tempurl%
\url{https://doi.org/10.1016/j.scs.2019.101925}
\showDOI{\tempurl}


\bibitem[Choudhary and Gokhale(2016)]%
        {choudhary2016urban}
\bibfield{author}{\bibinfo{person}{Arti Choudhary} {and}
  \bibinfo{person}{Sharad Gokhale}.} \bibinfo{year}{2016}\natexlab{}.
\newblock \showarticletitle{Urban real-world driving traffic emissions during
  interruption and congestion}.
\newblock \bibinfo{journal}{\emph{Transportation Research Part D: Transport and
  Environment}}  \bibinfo{volume}{43} (\bibinfo{year}{2016}),
  \bibinfo{pages}{59--70}.
\newblock
\showISSN{1361-9209}
\urldef\tempurl%
\url{https://doi.org/10.1016/j.trd.2015.12.006}
\showDOI{\tempurl}


\bibitem[Clauset et~al\mbox{.}(2009)]%
        {clauset2009powerlaw}
\bibfield{author}{\bibinfo{person}{Aaron Clauset},
  \bibinfo{person}{Cosma~Rohilla Shalizi}, {and} \bibinfo{person}{M.~E.~J.
  Newman}.} \bibinfo{year}{2009}\natexlab{}.
\newblock \showarticletitle{Power-Law Distributions in Empirical Data}.
\newblock \bibinfo{journal}{\emph{SIAM Rev.}} \bibinfo{volume}{51},
  \bibinfo{number}{4} (\bibinfo{year}{2009}), \bibinfo{pages}{661--703}.
\newblock
\urldef\tempurl%
\url{https://doi.org/10.1137/070710111}
\showDOI{\tempurl}


\bibitem[Coutrot et~al\mbox{.}(2022)]%
        {coutrot2022entropy}
\bibfield{author}{\bibinfo{person}{Antoine Coutrot}, \bibinfo{person}{Ed
  Manley}, \bibinfo{person}{S. Goodroe}, \bibinfo{person}{C. Gahnstrom},
  \bibinfo{person}{Gabriele Filomena}, \bibinfo{person}{Demet Yesiltepe},
  \bibinfo{person}{Ruth Dalton}, \bibinfo{person}{Jan Wiener},
  \bibinfo{person}{C. Hölscher}, \bibinfo{person}{Michael Hornberger}, {and}
  \bibinfo{person}{Hugo Spiers}.} \bibinfo{year}{2022}\natexlab{}.
\newblock \showarticletitle{Entropy of city street networks linked to future
  spatial navigation ability}.
\newblock \bibinfo{journal}{\emph{Nature}}  \bibinfo{volume}{604}
  (\bibinfo{date}{04} \bibinfo{year}{2022}).
\newblock
\urldef\tempurl%
\url{https://doi.org/10.1038/s41586-022-04486-7}
\showDOI{\tempurl}


\bibitem[{De Baets} et~al\mbox{.}(2014)]%
        {debaets2014route}
\bibfield{author}{\bibinfo{person}{Koen {De Baets}}, \bibinfo{person}{Sven
  Vlassenroot}, \bibinfo{person}{Kobe Boussauw}, \bibinfo{person}{Dirk
  Lauwers}, \bibinfo{person}{Georges Allaert}, {and} \bibinfo{person}{Philippe
  {De Maeyer}}.} \bibinfo{year}{2014}\natexlab{}.
\newblock \showarticletitle{Route choice and residential environment:
  introducing liveability requirements in navigation systems in Flanders}.
\newblock \bibinfo{journal}{\emph{Journal of Transport Geography}}
  \bibinfo{volume}{37} (\bibinfo{year}{2014}), \bibinfo{pages}{19--27}.
\newblock
\showISSN{0966-6923}
\urldef\tempurl%
\url{https://doi.org/10.1016/j.jtrangeo.2014.04.005}
\showDOI{\tempurl}


\bibitem[deSouza et~al\mbox{.}(2020)]%
        {desouza2020}
\bibfield{author}{\bibinfo{person}{Priyanka deSouza}, \bibinfo{person}{Amin
  Anjomshoaa}, \bibinfo{person}{Fabio Duarte}, \bibinfo{person}{Ralph Kahn},
  \bibinfo{person}{Prashant Kumar}, {and} \bibinfo{person}{Carlo Ratti}.}
  \bibinfo{year}{2020}\natexlab{}.
\newblock \showarticletitle{Air quality monitoring using mobile low-cost
  sensors mounted on trash-trucks: Methods development and lessons learned}.
\newblock \bibinfo{journal}{\emph{Sustainable Cities and Society}}
  \bibinfo{volume}{60} (\bibinfo{year}{2020}), \bibinfo{pages}{102239}.
\newblock
\showISSN{2210-6707}
\urldef\tempurl%
\url{https://doi.org/10.1016/j.scs.2020.102239}
\showDOI{\tempurl}


\bibitem[Ericsson et~al\mbox{.}(2006)]%
        {ericsson2006optimizing}
\bibfield{author}{\bibinfo{person}{Eva Ericsson}, \bibinfo{person}{Hanna
  Larsson}, {and} \bibinfo{person}{Karin Brundell-Freij}.}
  \bibinfo{year}{2006}\natexlab{}.
\newblock \showarticletitle{Optimizing route choice for lowest fuel consumption
  – Potential effects of a new driver support tool}.
\newblock \bibinfo{journal}{\emph{Transportation Research Part C: Emerging
  Technologies}} \bibinfo{volume}{14}, \bibinfo{number}{6}
  (\bibinfo{year}{2006}), \bibinfo{pages}{369--383}.
\newblock
\showISSN{0968-090X}
\urldef\tempurl%
\url{https://doi.org/10.1016/j.trc.2006.10.001}
\showDOI{\tempurl}


\bibitem[Ferreira et~al\mbox{.}(2015)]%
        {ferreira2015impact}
\bibfield{author}{\bibinfo{person}{João~C. Ferreira}, \bibinfo{person}{José
  de Almeida}, {and} \bibinfo{person}{Alberto~Rodrigues da Silva}.}
  \bibinfo{year}{2015}\natexlab{}.
\newblock \showarticletitle{The Impact of Driving Styles on Fuel Consumption: A
  Data-Warehouse-and-Data-Mining-Based Discovery Process}.
\newblock \bibinfo{journal}{\emph{IEEE Transactions on Intelligent
  Transportation Systems}} \bibinfo{volume}{16}, \bibinfo{number}{5}
  (\bibinfo{year}{2015}), \bibinfo{pages}{2653--2662}.
\newblock
\urldef\tempurl%
\url{https://doi.org/10.1109/TITS.2015.2414663}
\showDOI{\tempurl}


\bibitem[Fisher(2022)]%
        {fisher2022algorithms}
\bibfield{author}{\bibinfo{person}{Eran Fisher}.}
  \bibinfo{year}{2022}\natexlab{}.
\newblock \showarticletitle{Do algorithms have a right to the city? Waze and
  algorithmic spatiality}.
\newblock \bibinfo{journal}{\emph{Cultural Studies}} \bibinfo{volume}{36},
  \bibinfo{number}{1} (\bibinfo{year}{2022}), \bibinfo{pages}{74--95}.
\newblock
\urldef\tempurl%
\url{https://doi.org/10.1080/09502386.2020.1755711}
\showDOI{\tempurl}


\bibitem[Foderaro(2018)]%
        {leonia}
\bibfield{author}{\bibinfo{person}{Lisa~W. Foderaro}.}
  \bibinfo{year}{2018}\natexlab{}.
\newblock \bibinfo{booktitle}{\emph{New Jersey Town Aims to Keep App-Guided
  Outsiders Off Its Streets}}.
\newblock
\urldef\tempurl%
\url{https://www.nytimes.com/2018/01/22/nyregion/leonia-gps-navigation-apps.html}
\showURL{%
\tempurl}


\bibitem[Foo et~al\mbox{.}(2005)]%
        {foo2005do}
\bibfield{author}{\bibinfo{person}{Patrick Foo}, \bibinfo{person}{William
  Warren}, \bibinfo{person}{Andrew Duchon}, {and} \bibinfo{person}{Michael
  Tarr}.} \bibinfo{year}{2005}\natexlab{}.
\newblock \showarticletitle{Do Humans Integrate Routes Into a Cognitive Map?
  Map- Versus Landmark-Based Navigation of Novel Shortcuts.}
\newblock \bibinfo{journal}{\emph{Journal of experimental psychology. Learning,
  memory, and cognition}}  \bibinfo{volume}{31} (\bibinfo{date}{04}
  \bibinfo{year}{2005}), \bibinfo{pages}{195--215}.
\newblock
\urldef\tempurl%
\url{https://doi.org/10.1037/0278-7393.31.2.195}
\showDOI{\tempurl}


\bibitem[Gately et~al\mbox{.}(2017)]%
        {gately2017urban}
\bibfield{author}{\bibinfo{person}{Conor~K. Gately}, \bibinfo{person}{Lucy~R.
  Hutyra}, \bibinfo{person}{Scott Peterson}, {and} \bibinfo{person}{Ian {Sue
  Wing}}.} \bibinfo{year}{2017}\natexlab{}.
\newblock \showarticletitle{Urban emissions hotspots: Quantifying vehicle
  congestion and air pollution using mobile phone GPS data}.
\newblock \bibinfo{journal}{\emph{Environmental Pollution}}
  \bibinfo{volume}{229} (\bibinfo{year}{2017}), \bibinfo{pages}{496--504}.
\newblock
\showISSN{0269-7491}
\urldef\tempurl%
\url{https://doi.org/10.1016/j.envpol.2017.05.091}
\showDOI{\tempurl}


\bibitem[Hariharan and Toyama(2004)]%
        {Ramaswamy2004Project}
\bibfield{author}{\bibinfo{person}{Ramaswamy Hariharan} {and}
  \bibinfo{person}{Kentaro Toyama}.} \bibinfo{year}{2004}\natexlab{}.
\newblock \showarticletitle{Project Lachesis: Parsing and Modeling Location
  Histories}.
\newblock \bibinfo{journal}{\emph{Geographic Information Science. Volume 3234}}
   \bibinfo{volume}{3234}, \bibinfo{pages}{106--124}.
\newblock
\showISBNx{978-3-540-23558-3}
\urldef\tempurl%
\url{https://doi.org/10.1007/978-3-540-30231-5_8}
\showDOI{\tempurl}


\bibitem[i~Diao and Joseph~Ferreira(2014)]%
        {diao2014vehicle}
\bibfield{author}{\bibinfo{person}{M i Diao} {and} \bibinfo{person}{Jr
  Joseph~Ferreira}.} \bibinfo{year}{2014}\natexlab{}.
\newblock \showarticletitle{Vehicle Miles Traveled and the Built Environment:
  Evidence from Vehicle Safety Inspection Data}.
\newblock \bibinfo{journal}{\emph{Environment and Planning A: Economy and
  Space}} \bibinfo{volume}{46}, \bibinfo{number}{12} (\bibinfo{year}{2014}),
  \bibinfo{pages}{2991--3009}.
\newblock
\urldef\tempurl%
\url{https://doi.org/10.1068/a140039p}
\showDOI{\tempurl}


\bibitem[INFRAS(2013)]%
        {infras2013handbook}
\bibfield{author}{\bibinfo{person}{INFRAS}.} \bibinfo{year}{2013}\natexlab{}.
\newblock \bibinfo{title}{Handbuch für Emissionsfaktoren}.
\newblock \bibinfo{howpublished}{\url{http://www.hbefa.net/}}.
\newblock


\bibitem[Jensen et~al\mbox{.}(2020)]%
        {jensen2020route}
\bibfield{author}{\bibinfo{person}{Anders~F. Jensen},
  \bibinfo{person}{Thomas~K. Rasmussen}, {and} \bibinfo{person}{Carlo~G.
  Prato}.} \bibinfo{year}{2020}\natexlab{}.
\newblock \showarticletitle{A Route Choice Model for Capturing Driver
  Preferences When Driving Electric and Conventional Vehicles}.
\newblock \bibinfo{journal}{\emph{Sustainability}} \bibinfo{volume}{12},
  \bibinfo{number}{3} (\bibinfo{year}{2020}).
\newblock
\showISSN{2071-1050}
\urldef\tempurl%
\url{https://doi.org/10.3390/su12031149}
\showDOI{\tempurl}


\bibitem[Kancharla and Ramadurai(2018)]%
        {kancharla2018incorporating}
\bibfield{author}{\bibinfo{person}{Surendra~Reddy Kancharla} {and}
  \bibinfo{person}{Gitakrishnan Ramadurai}.} \bibinfo{year}{2018}\natexlab{}.
\newblock \showarticletitle{Incorporating driving cycle based fuel consumption
  estimation in green vehicle routing problems}.
\newblock \bibinfo{journal}{\emph{Sustainable Cities and Society}}
  \bibinfo{volume}{40} (\bibinfo{year}{2018}), \bibinfo{pages}{214--221}.
\newblock
\showISSN{2210-6707}
\urldef\tempurl%
\url{https://doi.org/10.1016/j.scs.2018.04.016}
\showDOI{\tempurl}


\bibitem[Krajzewicz et~al\mbox{.}(2015)]%
        {krajzewicz2015second}
\bibfield{author}{\bibinfo{person}{Daniel Krajzewicz}, \bibinfo{person}{Michael
  Behrisch}, \bibinfo{person}{Peter Wagner}, \bibinfo{person}{Raphael Luz},
  {and} \bibinfo{person}{Mario Krumnow}.} \bibinfo{year}{2015}\natexlab{}.
\newblock \showarticletitle{Second Generation of Pollutant Emission Models for
  SUMO}. In \bibinfo{booktitle}{\emph{Modeling Mobility with Open Data}},
  \bibfield{editor}{\bibinfo{person}{Michael Behrisch} {and}
  \bibinfo{person}{Melanie Weber}} (Eds.). \bibinfo{publisher}{Springer
  International Publishing}, \bibinfo{address}{Cham},
  \bibinfo{pages}{203--221}.
\newblock


\bibitem[Krajzewicz et~al\mbox{.}(2005)]%
        {krajzewicz2005simulation}
\bibfield{author}{\bibinfo{person}{Daniel Krajzewicz}, \bibinfo{person}{Elmar
  Brockfeld}, \bibinfo{person}{Jürgen Mikat}, \bibinfo{person}{Julia Ringel},
  \bibinfo{person}{Christian Feld}, \bibinfo{person}{Wolfram Tuchscheerer},
  \bibinfo{person}{Peter Wagner}, {and} \bibinfo{person}{Richard Woesler}.}
  \bibinfo{year}{2005}\natexlab{}.
\newblock \showarticletitle{Simulation of modern Traffic Lights Control Systems
  using the open source Traffic Simulation SUMO}. \bibinfo{pages}{299--302}.
\newblock
\showISBNx{90-77381-18-X}


\bibitem[Krajzewicz et~al\mbox{.}(2012)]%
        {krajzewicz2012recent}
\bibfield{author}{\bibinfo{person}{Daniel Krajzewicz}, \bibinfo{person}{Jakob
  Erdmann}, \bibinfo{person}{Michael Behrisch}, {and} \bibinfo{person}{Laura
  Bieker}.} \bibinfo{year}{2012}\natexlab{}.
\newblock \showarticletitle{Recent development and applications of
  SUMO-Simulation of Urban MObility}.
\newblock \bibinfo{journal}{\emph{International journal on advances in systems
  and measurements}} \bibinfo{volume}{5}, \bibinfo{number}{3\&4}
  (\bibinfo{year}{2012}).
\newblock


\bibitem[Lima et~al\mbox{.}(2016)]%
        {lima2016understanding}
\bibfield{author}{\bibinfo{person}{Antonio Lima}, \bibinfo{person}{Rade
  Stanojevic}, \bibinfo{person}{Dina Papagiannaki}, \bibinfo{person}{Pablo
  Rodriguez}, {and} \bibinfo{person}{Marta~C. González}.}
  \bibinfo{year}{2016}\natexlab{}.
\newblock \showarticletitle{Understanding individual routing behaviour}.
\newblock \bibinfo{journal}{\emph{Journal of The Royal Society Interface}}
  \bibinfo{volume}{13}, \bibinfo{number}{116} (\bibinfo{year}{2016}),
  \bibinfo{pages}{20160021}.
\newblock
\urldef\tempurl%
\url{https://doi.org/10.1098/rsif.2016.0021}
\showDOI{\tempurl}
\showeprint{https://royalsocietypublishing.org/doi/pdf/10.1098/rsif.2016.0021}


\bibitem[Liu et~al\mbox{.}(2019)]%
        {liu2019}
\bibfield{author}{\bibinfo{person}{Jielun Liu}, \bibinfo{person}{Ke Han},
  \bibinfo{person}{Xiqun~(Michael) Chen}, {and} \bibinfo{person}{Ghim~Ping
  Ong}.} \bibinfo{year}{2019}\natexlab{}.
\newblock \showarticletitle{Spatial-temporal inference of urban traffic
  emissions based on taxi trajectories and multi-source urban data}.
\newblock \bibinfo{journal}{\emph{Transportation Research Part C: Emerging
  Technologies}}  \bibinfo{volume}{106} (\bibinfo{year}{2019}),
  \bibinfo{pages}{145 -- 165}.
\newblock
\showISSN{0968-090X}
\urldef\tempurl%
\url{https://doi.org/10.1016/j.trc.2019.07.005}
\showDOI{\tempurl}


\bibitem[Lopez et~al\mbox{.}(2018)]%
        {Microscopic2018Lopez}
\bibfield{author}{\bibinfo{person}{Pablo~Alvarez Lopez},
  \bibinfo{person}{Michael Behrisch}, \bibinfo{person}{Laura Bieker-Walz},
  \bibinfo{person}{Jakob Erdmann}, \bibinfo{person}{Yun-Pang Flötteröd},
  \bibinfo{person}{Robert Hilbrich}, \bibinfo{person}{Leonhard Lücken},
  \bibinfo{person}{Johannes Rummel}, \bibinfo{person}{Peter Wagner}, {and}
  \bibinfo{person}{Evamarie Wiessner}.} \bibinfo{year}{2018}\natexlab{}.
\newblock \showarticletitle{Microscopic Traffic Simulation using SUMO}. In
  \bibinfo{booktitle}{\emph{2018 21st International Conference on Intelligent
  Transportation Systems (ITSC)}}. \bibinfo{pages}{2575--2582}.
\newblock
\urldef\tempurl%
\url{https://doi.org/10.1109/ITSC.2018.8569938}
\showDOI{\tempurl}


\bibitem[Luca et~al\mbox{.}(2021)]%
        {luca2020deep}
\bibfield{author}{\bibinfo{person}{Massimiliano Luca}, \bibinfo{person}{Gianni
  Barlacchi}, \bibinfo{person}{Bruno Lepri}, {and} \bibinfo{person}{Luca
  Pappalardo}.} \bibinfo{year}{2021}\natexlab{}.
\newblock \showarticletitle{A Survey on Deep Learning for Human Mobility}.
\newblock \bibinfo{journal}{\emph{ACM Comput. Surv.}} \bibinfo{volume}{55},
  \bibinfo{number}{1}, Article \bibinfo{articleno}{7} (\bibinfo{date}{nov}
  \bibinfo{year}{2021}), \bibinfo{numpages}{44}~pages.
\newblock
\showISSN{0360-0300}
\urldef\tempurl%
\url{https://doi.org/10.1145/3485125}
\showDOI{\tempurl}


\bibitem[Luján et~al\mbox{.}(2018)]%
        {lujan2018}
\bibfield{author}{\bibinfo{person}{José~M. Luján}, \bibinfo{person}{Vicente
  Bermúdez}, \bibinfo{person}{Vicente Dolz}, {and} \bibinfo{person}{Javier
  Monsalve-Serrano}.} \bibinfo{year}{2018}\natexlab{}.
\newblock \showarticletitle{An assessment of the real-world driving gaseous
  emissions from a Euro 6 light-duty diesel vehicle using a portable emissions
  measurement system (PEMS)}.
\newblock \bibinfo{journal}{\emph{Atmospheric Environment}}
  \bibinfo{volume}{174} (\bibinfo{year}{2018}), \bibinfo{pages}{112 -- 121}.
\newblock
\showISSN{1352-2310}
\urldef\tempurl%
\url{https://doi.org/10.1016/j.atmosenv.2017.11.056}
\showDOI{\tempurl}


\bibitem[Macfarlane(2019)]%
        {macfarlane2019when}
\bibfield{author}{\bibinfo{person}{Jane Macfarlane}.}
  \bibinfo{year}{2019}\natexlab{}.
\newblock \showarticletitle{When apps rule the road: The proliferation of
  navigation apps is causing traffic chaos. It's time to restore order}.
\newblock \bibinfo{journal}{\emph{IEEE Spectrum}} \bibinfo{volume}{56},
  \bibinfo{number}{10} (\bibinfo{year}{2019}), \bibinfo{pages}{22--27}.
\newblock
\urldef\tempurl%
\url{https://doi.org/10.1109/MSPEC.2019.8847586}
\showDOI{\tempurl}


\bibitem[Malik et~al\mbox{.}(2019)]%
        {malik2019evaluation}
\bibfield{author}{\bibinfo{person}{Fehda Malik}, \bibinfo{person}{Hasan~Ali
  Khattak}, {and} \bibinfo{person}{Munam Ali~Shah}.}
  \bibinfo{year}{2019}\natexlab{}.
\newblock \showarticletitle{Evaluation of the Impact of Traffic Congestion
  Based on SUMO}. In \bibinfo{booktitle}{\emph{2019 25th International
  Conference on Automation and Computing (ICAC)}}. \bibinfo{pages}{1--5}.
\newblock
\urldef\tempurl%
\url{https://doi.org/10.23919/IConAC.2019.8895120}
\showDOI{\tempurl}


\bibitem[Manley et~al\mbox{.}(2015)]%
        {manley2015shortest}
\bibfield{author}{\bibinfo{person}{E.J. Manley}, \bibinfo{person}{J.D.
  Addison}, {and} \bibinfo{person}{T. Cheng}.} \bibinfo{year}{2015}\natexlab{}.
\newblock \showarticletitle{Shortest path or anchor-based route choice: a
  large-scale empirical analysis of minicab routing in London}.
\newblock \bibinfo{journal}{\emph{Journal of Transport Geography}}
  \bibinfo{volume}{43} (\bibinfo{year}{2015}), \bibinfo{pages}{123--139}.
\newblock
\showISSN{0966-6923}
\urldef\tempurl%
\url{https://doi.org/10.1016/j.jtrangeo.2015.01.006}
\showDOI{\tempurl}


\bibitem[Mehrvarz et~al\mbox{.}(2020)]%
        {mehrvarz2020optimal}
\bibfield{author}{\bibinfo{person}{Nassim Mehrvarz}, \bibinfo{person}{Zhilin
  Ye}, \bibinfo{person}{Khalegh Barati}, {and} \bibinfo{person}{Xuesong Shen}.}
  \bibinfo{year}{2020}\natexlab{}.
\newblock \showarticletitle{Optimal Travel Routes of On-road Vehicles
  Considering Sustainability}. In \bibinfo{booktitle}{\emph{Proceedings of the
  37th International Symposium on Automation and Robotics in Construction
  (ISARC)}}. \bibinfo{publisher}{International Association for Automation and
  Robotics in Construction (IAARC)}, \bibinfo{address}{Kitakyushu, Japan},
  \bibinfo{pages}{507--513}.
\newblock
\showISBNx{978-952-94-3634-7}
\showISSN{2413-5844}
\urldef\tempurl%
\url{https://doi.org/10.22260/ISARC2020/0070}
\showDOI{\tempurl}


\bibitem[Norman et~al\mbox{.}(2005)]%
        {norman2005perception}
\bibfield{author}{\bibinfo{person}{J~Farley Norman}, \bibinfo{person}{Charles~E
  Crabtree}, \bibinfo{person}{Anna~Marie Clayton}, {and}
  \bibinfo{person}{Hideko~F Norman}.} \bibinfo{year}{2005}\natexlab{}.
\newblock \showarticletitle{The Perception of Distances and Spatial
  Relationships in Natural Outdoor Environments}.
\newblock \bibinfo{journal}{\emph{Perception}} \bibinfo{volume}{34},
  \bibinfo{number}{11} (\bibinfo{year}{2005}), \bibinfo{pages}{1315--1324}.
\newblock
\urldef\tempurl%
\url{https://doi.org/10.1068/p5304}
\showDOI{\tempurl}
\showeprint{ttps://doi.org/10.1068/p5304}
\newblock
\shownote{PMID: 16355740}.


\bibitem[Nyhan et~al\mbox{.}(2016)]%
        {nyhan2016}
\bibfield{author}{\bibinfo{person}{Marguerite Nyhan},
  \bibinfo{person}{Stanislav Sobolevsky}, \bibinfo{person}{Chaogui Kang},
  \bibinfo{person}{Prudence Robinson}, \bibinfo{person}{Andrea Corti},
  \bibinfo{person}{Michael Szell}, \bibinfo{person}{David Streets},
  \bibinfo{person}{Zifeng Lu}, \bibinfo{person}{Rex Britter},
  \bibinfo{person}{Steven~R.H. Barrett}, {and} \bibinfo{person}{Carlo Ratti}.}
  \bibinfo{year}{2016}\natexlab{}.
\newblock \showarticletitle{Predicting vehicular emissions in high spatial
  resolution using pervasively measured transportation data and microscopic
  emissions model}.
\newblock \bibinfo{journal}{\emph{Atmospheric Environment}}
  \bibinfo{volume}{140} (\bibinfo{year}{2016}), \bibinfo{pages}{352 -- 363}.
\newblock
\showISSN{1352-2310}
\urldef\tempurl%
\url{https://doi.org/10.1016/j.atmosenv.2016.06.018}
\showDOI{\tempurl}


\bibitem[Pappalardo et~al\mbox{.}(2013)]%
        {pappalardo2013understanding}
\bibfield{author}{\bibinfo{person}{L. Pappalardo}, \bibinfo{person}{S.
  Rinzivillo}, \bibinfo{person}{Z. Qu}, \bibinfo{person}{D. Pedreschi}, {and}
  \bibinfo{person}{F. Giannotti}.} \bibinfo{year}{2013}\natexlab{}.
\newblock \showarticletitle{{Understanding the patterns of car travel}}.
\newblock \bibinfo{journal}{\emph{European Physical Journal: Special Topics}}
  \bibinfo{volume}{215}, \bibinfo{number}{1} (\bibinfo{year}{2013}).
\newblock
\showISSN{19516355}
\urldef\tempurl%
\url{https://doi.org/10.1140/epjst/e2013-01715-5}
\showDOI{\tempurl}


\bibitem[Pappalardo et~al\mbox{.}(2019)]%
        {scikitmob}
\bibfield{author}{\bibinfo{person}{Luca Pappalardo}, \bibinfo{person}{Filippo
  Simini}, \bibinfo{person}{Gianni Barlacchi}, {and} \bibinfo{person}{Roberto
  Pellungrini}.} \bibinfo{year}{2019}\natexlab{}.
\newblock \bibinfo{title}{Scikit-mobility: a Python library for the analysis,
  generation and risk assessment of mobility data}.
\newblock
\newblock
\urldef\tempurl%
\url{https://doi.org/10.48550/ARXIV.1907.07062}
\showDOI{\tempurl}


\bibitem[Perez-Prada et~al\mbox{.}(2017)]%
        {perezprada2017managing}
\bibfield{author}{\bibinfo{person}{Fiamma Perez-Prada},
  \bibinfo{person}{Andrés Monzón}, {and} \bibinfo{person}{Cristina Valdés}.}
  \bibinfo{year}{2017}\natexlab{}.
\newblock \showarticletitle{Managing Traffic Flows for Cleaner Cities: The Role
  of Green Navigation Systems}.
\newblock \bibinfo{journal}{\emph{Energies}}  \bibinfo{volume}{10}
  (\bibinfo{date}{06} \bibinfo{year}{2017}), \bibinfo{pages}{791}.
\newblock
\urldef\tempurl%
\url{https://doi.org/10.3390/en10060791}
\showDOI{\tempurl}


\bibitem[Rahman and Idris(2017)]%
        {rahman2017tribute}
\bibfield{author}{\bibinfo{person}{Md.~Nobinur Rahman} {and}
  \bibinfo{person}{Ahmed~O. Idris}.} \bibinfo{year}{2017}\natexlab{}.
\newblock \showarticletitle{TRIBUTE: Trip-based urban transportation emissions
  model for municipalities}.
\newblock \bibinfo{journal}{\emph{International Journal of Sustainable
  Transportation}} \bibinfo{volume}{11}, \bibinfo{number}{7}
  (\bibinfo{year}{2017}), \bibinfo{pages}{540--552}.
\newblock
\urldef\tempurl%
\url{https://doi.org/10.1080/15568318.2016.1278061}
\showDOI{\tempurl}


\bibitem[Reznik et~al\mbox{.}(2019)]%
        {reznik2018}
\bibfield{author}{\bibinfo{person}{Ariel Reznik}, \bibinfo{person}{Meidad
  Kissinger}, {and} \bibinfo{person}{Nurit Alfasi}.}
  \bibinfo{year}{2019}\natexlab{}.
\newblock \showarticletitle{Real-data-based high-resolution GHG emissions
  accounting of urban residents private transportation}.
\newblock \bibinfo{journal}{\emph{International Journal of Sustainable
  Transportation}} \bibinfo{volume}{13}, \bibinfo{number}{4}
  (\bibinfo{year}{2019}), \bibinfo{pages}{235--244}.
\newblock
\urldef\tempurl%
\url{https://doi.org/10.1080/15568318.2018.1459971}
\showDOI{\tempurl}


\bibitem[Samaras et~al\mbox{.}(2016)]%
        {samaras2016quantification}
\bibfield{author}{\bibinfo{person}{Zissis Samaras}, \bibinfo{person}{Leonidas
  Ntziachristos}, \bibinfo{person}{Silvana Toffolo}, \bibinfo{person}{Giorgio
  Magra}, \bibinfo{person}{Alvaro Garcia-Castro}, \bibinfo{person}{Cristina
  Valdes}, \bibinfo{person}{Christian Vock}, {and} \bibinfo{person}{Werner
  Maier}.} \bibinfo{year}{2016}\natexlab{}.
\newblock \showarticletitle{Quantification of the Effect of ITS on CO2
  Emissions from Road Transportation}.
\newblock \bibinfo{journal}{\emph{Transportation Research Procedia}}
  \bibinfo{volume}{14} (\bibinfo{year}{2016}), \bibinfo{pages}{3139--3148}.
\newblock
\showISSN{2352-1465}
\urldef\tempurl%
\url{https://doi.org/10.1016/j.trpro.2016.05.254}
\showDOI{\tempurl}
\newblock
\shownote{Transport Research Arena TRA2016}.


\bibitem[Seele et~al\mbox{.}(2012)]%
        {Seele2012Cognitive}
\bibfield{author}{\bibinfo{person}{Sven Seele}, \bibinfo{person}{Thomas
  Dettmar}, \bibinfo{person}{Rainer Herpers}, \bibinfo{person}{Christian
  Bauckhage}, {and} \bibinfo{person}{Peter Becker}.}
  \bibinfo{year}{2012}\natexlab{}.
\newblock \showarticletitle{Cognitive Aspects of Traffic Simulations in Virtual
  Environments}.
\newblock \bibinfo{journal}{\emph{Simul. Notes Eur.}}  \bibinfo{volume}{22}
  (\bibinfo{year}{2012}), \bibinfo{pages}{83--88}.
\newblock


\bibitem[Siuhi and Mwakalonge(2016)]%
        {siuhi2016opportunities}
\bibfield{author}{\bibinfo{person}{Saidi Siuhi} {and} \bibinfo{person}{Judith
  Mwakalonge}.} \bibinfo{year}{2016}\natexlab{}.
\newblock \showarticletitle{Opportunities and challenges of smart mobile
  applications in transportation}.
\newblock \bibinfo{journal}{\emph{Journal of Traffic and Transportation
  Engineering (English Edition)}} \bibinfo{volume}{3}, \bibinfo{number}{6}
  (\bibinfo{year}{2016}), \bibinfo{pages}{582--592}.
\newblock
\showISSN{2095-7564}
\urldef\tempurl%
\url{https://doi.org/10.1016/j.jtte.2016.11.001}
\showDOI{\tempurl}


\bibitem[Sui et~al\mbox{.}(2019)]%
        {sui2019gps}
\bibfield{author}{\bibinfo{person}{Yi Sui}, \bibinfo{person}{Haoran Zhang},
  \bibinfo{person}{Xuan Song}, \bibinfo{person}{Fengjing Shao},
  \bibinfo{person}{Xiang Yu}, \bibinfo{person}{Ryosuke Shibasaki},
  \bibinfo{person}{Rechengcheng Sun}, \bibinfo{person}{Meng Yuan},
  \bibinfo{person}{Changying Wang}, \bibinfo{person}{Shujing Li}, {and}
  \bibinfo{person}{Yao Li}.} \bibinfo{year}{2019}\natexlab{}.
\newblock \showarticletitle{GPS data in urban online ride-hailing: A
  comparative analysis on fuel consumption and emissions}.
\newblock \bibinfo{journal}{\emph{Journal of Cleaner Production}}
  \bibinfo{volume}{227} (\bibinfo{year}{2019}), \bibinfo{pages}{495 -- 505}.
\newblock
\showISSN{0959-6526}
\urldef\tempurl%
\url{https://doi.org/10.1016/j.jclepro.2019.04.159}
\showDOI{\tempurl}


\bibitem[{United Nations General Assembly}(2015)]%
        {assembly2015sustainable}
\bibfield{author}{\bibinfo{person}{{United Nations General Assembly}}.}
  \bibinfo{year}{2015}\natexlab{}.
\newblock \bibinfo{booktitle}{\emph{Transforming our world: the 2030 Agenda for
  Sustainable Development}}.
\newblock \bibinfo{type}{{T}echnical {R}eport}.
\newblock
\newblock
\shownote{Accessed: 2021-02-23}.


\bibitem[Wang et~al\mbox{.}(2014)]%
        {wang2014using}
\bibfield{author}{\bibinfo{person}{Xiaoguang Wang}, \bibinfo{person}{Joe
  Grengs}, {and} \bibinfo{person}{Lidia Kostyniuk}.}
  \bibinfo{year}{2014}\natexlab{}.
\newblock \showarticletitle{Using a GPS Data Set to Examine the Effects of the
  Built Environment along Commuting Routes on Travel Outcomes}.
\newblock \bibinfo{journal}{\emph{Journal of Urban Planning and Development}}
  \bibinfo{volume}{140}, \bibinfo{number}{4} (\bibinfo{year}{2014}),
  \bibinfo{pages}{04014009}.
\newblock
\urldef\tempurl%
\url{https://doi.org/10.1061/(ASCE)UP.1943-5444.0000181}
\showDOI{\tempurl}


\bibitem[Xu et~al\mbox{.}(2021)]%
        {xu2021understanding}
\bibfield{author}{\bibinfo{person}{Yanyan Xu}, \bibinfo{person}{Riccardo
  Di~Clemente}, {and} \bibinfo{person}{Marta~C. Gonzalez}.}
  \bibinfo{year}{2021}\natexlab{}.
\newblock \showarticletitle{Understanding vehicular routing behavior with
  location-based service data}.
\newblock \bibinfo{journal}{\emph{EPJ Data Science}}  \bibinfo{volume}{10}
  (\bibinfo{date}{12} \bibinfo{year}{2021}).
\newblock
\urldef\tempurl%
\url{https://doi.org/10.1140/epjds/s13688-021-00267-w}
\showDOI{\tempurl}


\bibitem[Yu et~al\mbox{.}(2020)]%
        {yu2020mobile}
\bibfield{author}{\bibinfo{person}{Qing Yu}, \bibinfo{person}{Haoran Zhang},
  \bibinfo{person}{Weifeng Li}, \bibinfo{person}{Xuan Song},
  \bibinfo{person}{Dongyuan Yang}, {and} \bibinfo{person}{Ryosuke Shibasaki}.}
  \bibinfo{year}{2020}\natexlab{}.
\newblock \showarticletitle{Mobile phone GPS data in urban customized bus:
  Dynamic line design and emission reduction potentials analysis}.
\newblock \bibinfo{journal}{\emph{Journal of Cleaner Production}}
  \bibinfo{volume}{272} (\bibinfo{year}{2020}), \bibinfo{pages}{122471}.
\newblock
\showISSN{0959-6526}
\urldef\tempurl%
\url{https://doi.org/10.1016/j.jclepro.2020.122471}
\showDOI{\tempurl}


\bibitem[Zheng et~al\mbox{.}(2017)]%
        {zheng2017influence}
\bibfield{author}{\bibinfo{person}{Fangfang Zheng}, \bibinfo{person}{Jie Li},
  \bibinfo{person}{Henk {van Zuylen}}, {and} \bibinfo{person}{Chao Lu}.}
  \bibinfo{year}{2017}\natexlab{}.
\newblock \showarticletitle{Influence of driver characteristics on emissions
  and fuel consumption}.
\newblock \bibinfo{journal}{\emph{Transportation Research Procedia}}
  \bibinfo{volume}{27} (\bibinfo{year}{2017}), \bibinfo{pages}{624--631}.
\newblock
\showISSN{2352-1465}
\urldef\tempurl%
\url{https://doi.org/10.1016/j.trpro.2017.12.142}
\showDOI{\tempurl}
\newblock
\shownote{20th EURO Working Group on Transportation Meeting, EWGT 2017, 4-6
  September 2017, Budapest, Hungary}.


\bibitem[Zhu et~al\mbox{.}(2020)]%
        {zhu2020high}
\bibfield{author}{\bibinfo{person}{Sicong Zhu}, \bibinfo{person}{Inhi Kim},
  {and} \bibinfo{person}{Keechoo Choi}.} \bibinfo{year}{2020}\natexlab{}.
\newblock \showarticletitle{High-resolution simulation-based analysis of
  leading vehicle acceleration profiles at signalized intersections for
  emission modeling}.
\newblock \bibinfo{journal}{\emph{International Journal of Sustainable
  Transportation}} \bibinfo{volume}{0}, \bibinfo{number}{0}
  (\bibinfo{year}{2020}), \bibinfo{pages}{1--11}.
\newblock
\urldef\tempurl%
\url{https://doi.org/10.1080/15568318.2020.1792011}
\showDOI{\tempurl}


\bibitem[Zhu and Levinson(2015)]%
        {zhu2015do}
\bibfield{author}{\bibinfo{person}{Shanjiang Zhu} {and} \bibinfo{person}{David
  Levinson}.} \bibinfo{year}{2015}\natexlab{}.
\newblock \showarticletitle{Do People Use the Shortest Path? An Empirical Test
  of Wardrop’s First Principle}.
\newblock \bibinfo{journal}{\emph{PLOS ONE}} \bibinfo{volume}{10},
  \bibinfo{number}{8} (\bibinfo{date}{08} \bibinfo{year}{2015}),
  \bibinfo{pages}{1--18}.
\newblock
\urldef\tempurl%
\url{https://doi.org/10.1371/journal.pone.0134322}
\showDOI{\tempurl}


\bibitem[Zubillaga et~al\mbox{.}(2014)]%
        {zubillaga2014measuring}
\bibfield{author}{\bibinfo{person}{Dario Zubillaga}, \bibinfo{person}{Geovany
  Cruz}, \bibinfo{person}{Luis Aguilar}, \bibinfo{person}{Jorge Zapotecatl},
  \bibinfo{person}{Nelson Fernández}, \bibinfo{person}{Jose Aguilar},
  \bibinfo{person}{David Rosenblueth}, {and} \bibinfo{person}{Carlos
  Gershenson}.} \bibinfo{year}{2014}\natexlab{}.
\newblock \showarticletitle{Measuring the Complexity of Self-Organizing Traffic
  Lights}.
\newblock \bibinfo{journal}{\emph{Entropy}}  \bibinfo{volume}{16}
  (\bibinfo{date}{02} \bibinfo{year}{2014}).
\newblock
\urldef\tempurl%
\url{https://doi.org/10.3390/e16052384}
\showDOI{\tempurl}


\end{thebibliography}

\clearpage
\section*{\huge{Appendix}}
\renewcommand\thefigure{A.\arabic{figure}}
\setcounter{figure}{0}

\renewcommand\thetable{AT.\arabic{table}}
\setcounter{table}{0}

\appendix

\section{Curve Fitting}
\label{sec:fitting_distributions}

We fit the same distributions using a maximum-likelihood \cite{clauset2009powerlaw, alstott2013powerlaw}. 
In particular, we fit five models to the data -- power-law, truncated power-law, lognormal, exponential, and stretched exponential -- and compare pairwise their goodness-of-fit with a log-likelihood ratio test. 
We choose the best fitting model as the one that wins the highest number of comparisons.

\begin{figure}[hbt!]
    \centering
    \includegraphics[width=0.8\linewidth]{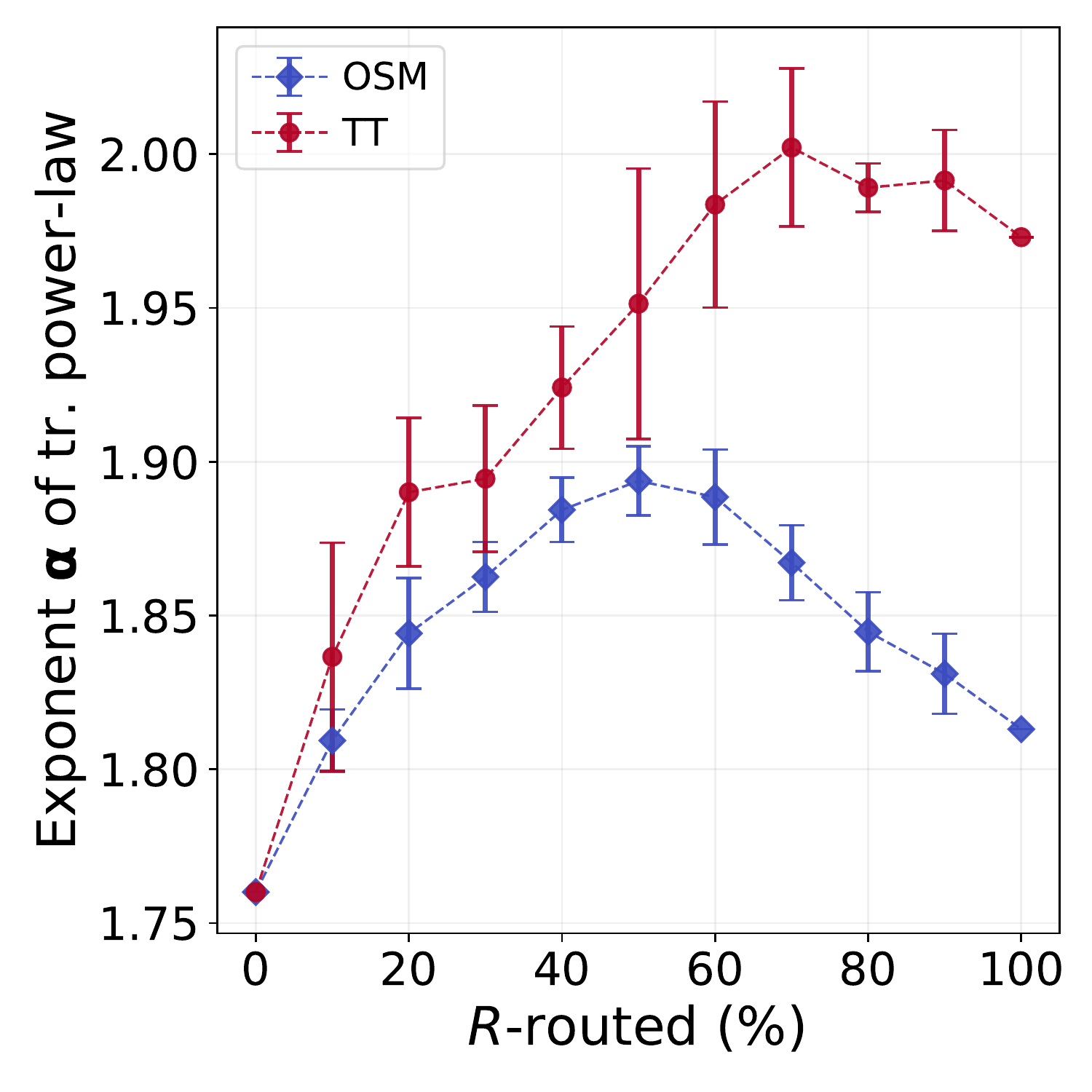}
    \caption{
    Exponent $\alpha$ of the truncated power-law fit to the CO$_2$ emissions distribution across the roads varying the percentage of $R$-routed vehicles, for OSM (blue) and TT (red).
    In the error bars, points indicate the average $\alpha$ over ten simulations with different $R$-routed vehicles chosen uniformly at random.
    Vertical bars indicate the standard deviation.
    }
    \label{fig:alpha}
\end{figure}

The best fit for the distributions is always a truncated power-law (essentially, a power-law with an exponential cutoff), with probability density function $p(x) \propto x^{-\alpha} e^{-\lambda x}$. 
Figure~\ref{fig:alpha} shows how the truncated power-law's exponent $\alpha$ varies with $\overline{D}_i^{(R)}$, $i = 0, \dots, 10$ and $R \in \{\text{OSM}, \text{TT}\}$.

\section{Traffic Calibration}
\label{sec:traf-cal}
To generate a plausible mobility demand reflecting real traffic patterns, we tune the number $N$ of vehicles in the simulation, the value of the parameter $w \in [1, +\infty)$ for DR, and the quantity $c$ of extra vehicles to insert during the simulation, where $c=none$ means no extra vehicles, $x_{\text{\footnotesize start}}$ means to add $N\frac{x_{\text{\footnotesize start}}}{100}$ extra vehicles at the start of the simulation; similarly, $x_{\text{\footnotesize end}}$ means to add $N\frac{x_{\text{\footnotesize end}}}{100}$ extra vehicles after the last trip's departure time (see Table \ref{tab:hyp_calibration} for details).

\begin{table}[htb!]
\centering
\begin{tabular}{@{}cll@{}}
\toprule
parameter & description & \multicolumn{1}{c}{values} \\ \midrule
$N$ & \footnotesize{number of vehicles} & \small{\{5000, 10000, 15000, 20000\}} \\
$w$ & \footnotesize{DR's randomisation parameter} & \small{\{1, 2.5, 5, 10, 15, 25\}} \\
$c$ & \footnotesize{configurations of extra vehicles} & \small \{none, $15_{start}$, $15_{start}$+$45_{end}$\} \\ \bottomrule
\end{tabular}
\caption{The parameters considered during the traffic calibration phase, a brief description, and the tuned values. }
\label{tab:hyp_calibration}
\end{table}

\begin{table}[htb!]
\centering
\begin{tabular}{@{}llll@{}}
\toprule
$C_{(N,w,c)}$ & JS & $|\Delta_{tt}|$ & teleports \\ \midrule
$C_{(15k,15,none)}$ & \textbf{0.232} & 57.4 & 3376.8 \\
$C_{(15k,25,15_{start})}$ & 0.233 & 201.34 & 5339.4 \\
\rowcolor{lightgray}
$C_{(15k,5,none)}$ & 0.234 & \textbf{32.94} & \textbf{3297.8} \\
$C_{(15k,5,15_{start})}$ & 0.234 & 198.46 & 5059.2 \\
$C_{(15k,25,none)}$ & 0.234 & 56.22 & 3518.8 \\ \bottomrule
\end{tabular}
\caption{Top five configurations sorted by JS. JS = Jensen-Shannon divergence, $|\Delta_{tt}|$ = average absolute travel time difference, teleports = number of teleports during the simulation. In gray, we highlight the selected configuration.
}
\label{tab:table_res_traffic_calibration}
\end{table}

\begin{figure}[hbt!]
    \centering
    \includegraphics[width=0.8\linewidth]{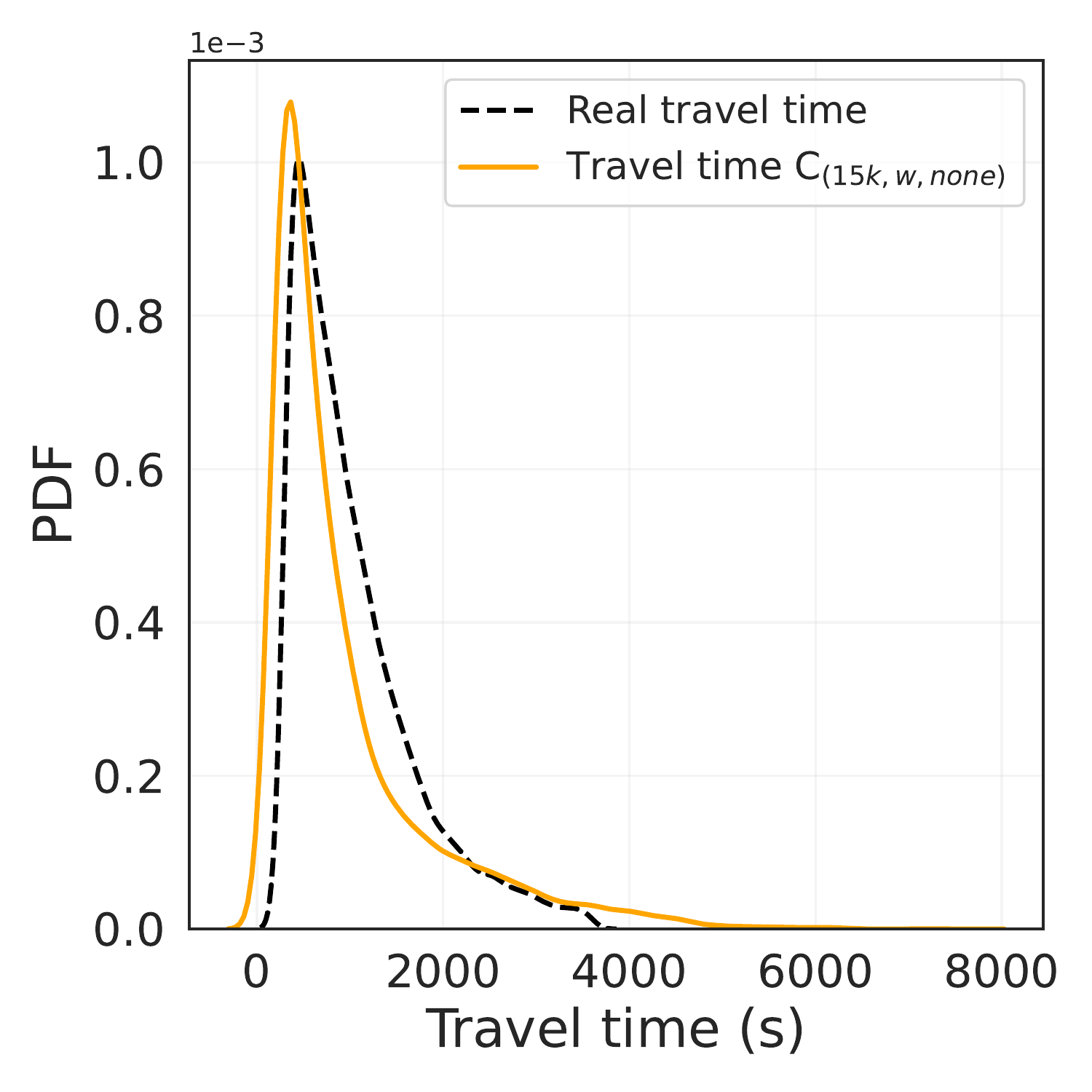}
    \caption{The kernel density estimation (KDE) of the probability distribution of the simulated vehicles' travel time considering a mobility demand obtained with the configuration $C_{(15k,5,none)}$ (orange solid line) and the distribution of the real vehicles' travel times (black dashed line).
    }
    \label{fig:kde_traffic_calibration}
\end{figure}

To assess the realism of the mobility demand $D$, we generate a multiset of routed paths $\overline{D}_0^{(R)}$ (no $R$-routed vehicles) and simulate the vehicular traffic using SUMO. Then, we use the Jensen–Shannon (JS) divergence \cite{luca2020deep} to compute the distance between the travel time distribution in real data and that of the simulated vehicles (see Figure \ref{fig:kde_traffic_calibration}).
The JS divergence ranges in $[0, 1]$ (the higher, the more similar the two distributions are) and is defined as \cite{luca2020deep}:
$$JS(P || Q) = \frac{1}{2} KL(P || M) + \frac{1}{2} KL(Q || M)$$
where $P$ and $Q$ are two density distributions, 
$M = \frac{1}{2} (P + Q)$, and $KL$ is the Kullback–Leibler divergence (KL), defined as:
$$KL(P||Q) = \sum_{x \in X} P(x) \log \left(\frac{P(x)}{Q(x)} \right)$$

We also assess the realism of the simulation using the absolute difference in seconds between the real and simulated average travel time $|\Delta_{tt}|$ and the number of teleports during the simulation. 
SUMO makes a teleport whenever a vehicle waits too long stuck in gridlock: the vehicle is teleported onto the next free edge on its path. 
We simulate each $\overline{D}_0^{(R)}$ five times and consider the average values of the above three metrics.

We run several configurations $C_{(N, w, c)}$ composing all combinations of $N$, $w$, and $c$ in Table \ref{tab:hyp_calibration}.
We select configuration $C_{(15k,5,\text{\footnotesize none)}}$, as it is the best  on two out three of the evaluation criteria: the JS difference wrt the minimum is only of $=0.002$, $|\Delta_{tt}| \approx 25$s, 3298 teleports, and $w=5$ (see Table \ref{tab:table_res_traffic_calibration}).


\section{Perturbation of the fastest path}
\label{sec:perturbation}
\begin{figure}[hbt!]
    \centering
    \includegraphics[width=0.7\linewidth]{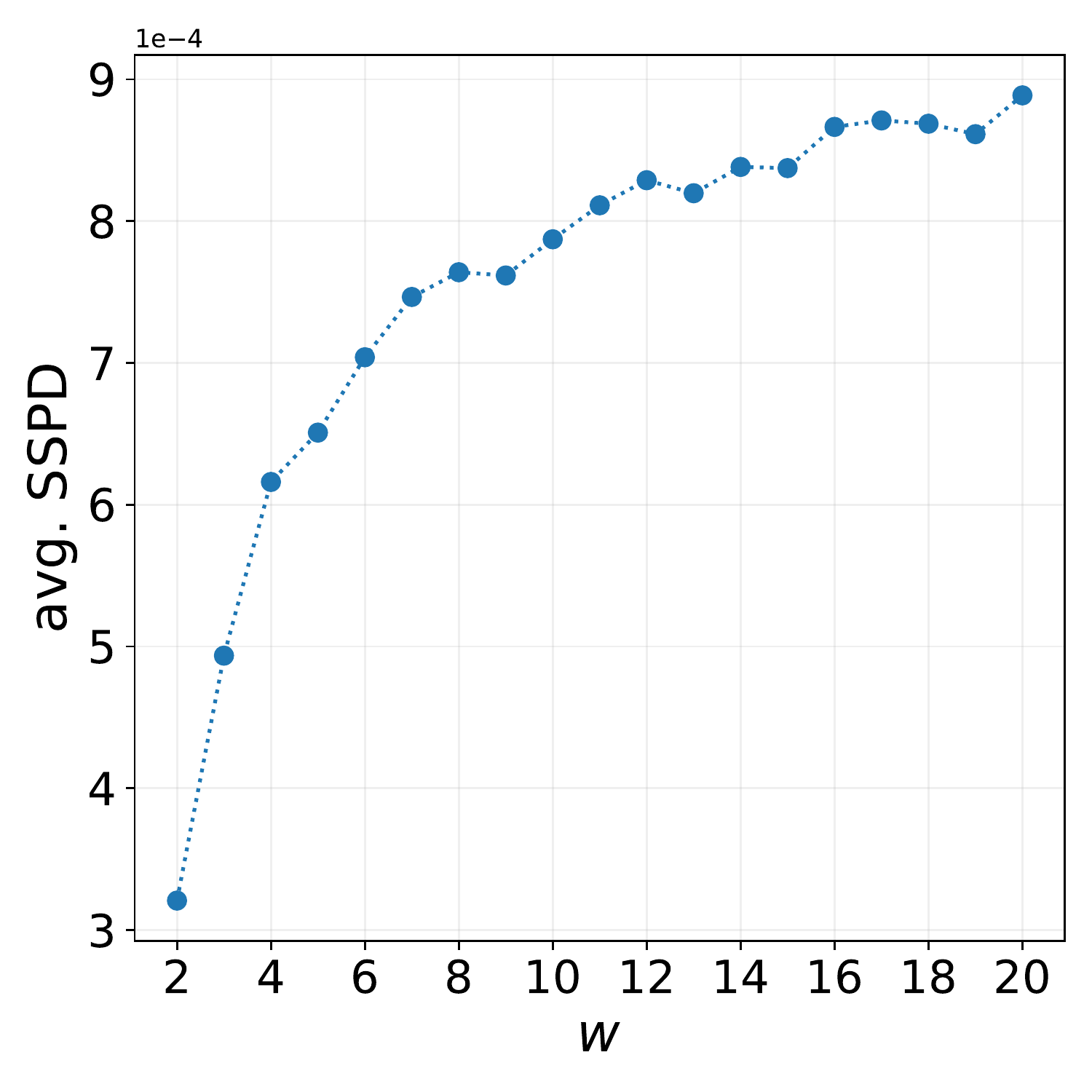}
    \caption{The average Symmetrized Segment-Path Distance (SSPD) computed between 15k perturbated paths (for different values of $w \in \{x \in \mathbb{N} | 2\leq x \leq 20\}$) and the fastest path ($w=1$). The higher $w$, the higher the a path's perturbation.}
    \label{fig:sspd_vs_w}
\end{figure}

To model the driving behavior of vehicles that are not $R$-routed (i.e., they do not follow any navigation apps' suggestion) and the imperfection and non-rationality of human drivers \cite{Seele2012Cognitive}, we perturb the fastest path between an origin and destination pair.
DR allows perturbing the fastest path using a randomisation parameter $w \in [1, +\infty)$, where $w=1$ means no randomisation (i.e., the fastest path), and the higher $w$, the more randomly perturbed the fastest path is.
To confirm that the perturbation of a path from an origin to a destination grows with $w$, we take randomly 15,000 origin-destination pairs computing the path suggested by DR for different values of $w \in \{x \in \mathbb{N} | 1\leq x \leq 20\}$. Next, we measure the perturbation of a path as the Symmetrized Segment-Path Distance (SSPD) \cite{Review2015Besse} between the perturbated paths ($w>1$) and the fastest path ($w=1$).
Figure \ref{fig:sspd_vs_w} shows that increasing the value of $w$ results in paths that have a greater average SSPD with respect to the fastest path, and hence that the perturbation grows with increasing $w$.

\section{Spatial distribution of emissions}
\label{sec:emissions_diff}
\begin{figure}[htb!]
    \centering
    \subfigure[]{
    \includegraphics[width=0.7\columnwidth]{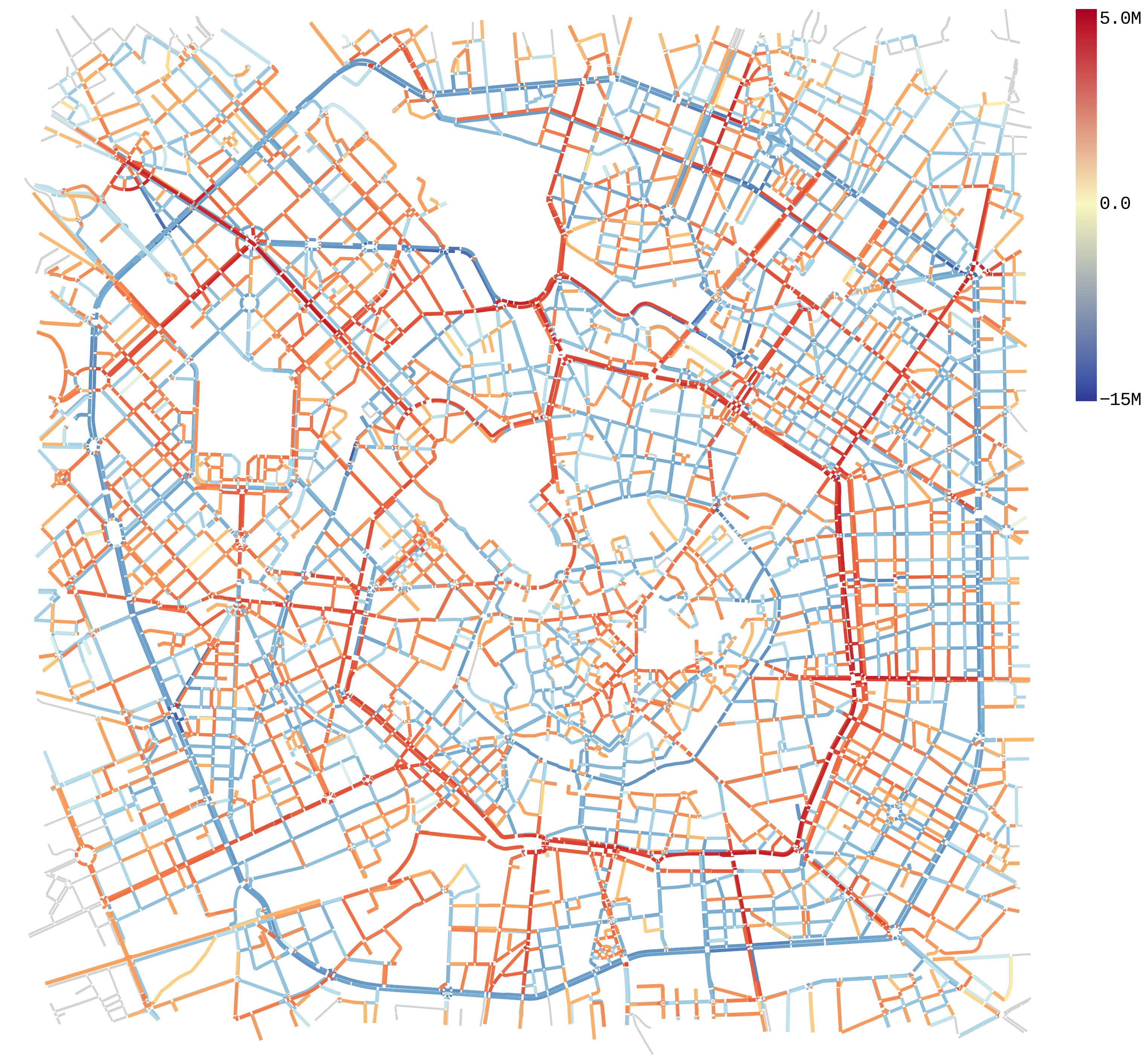}}
    \subfigure[]{\includegraphics[width=0.7\columnwidth]{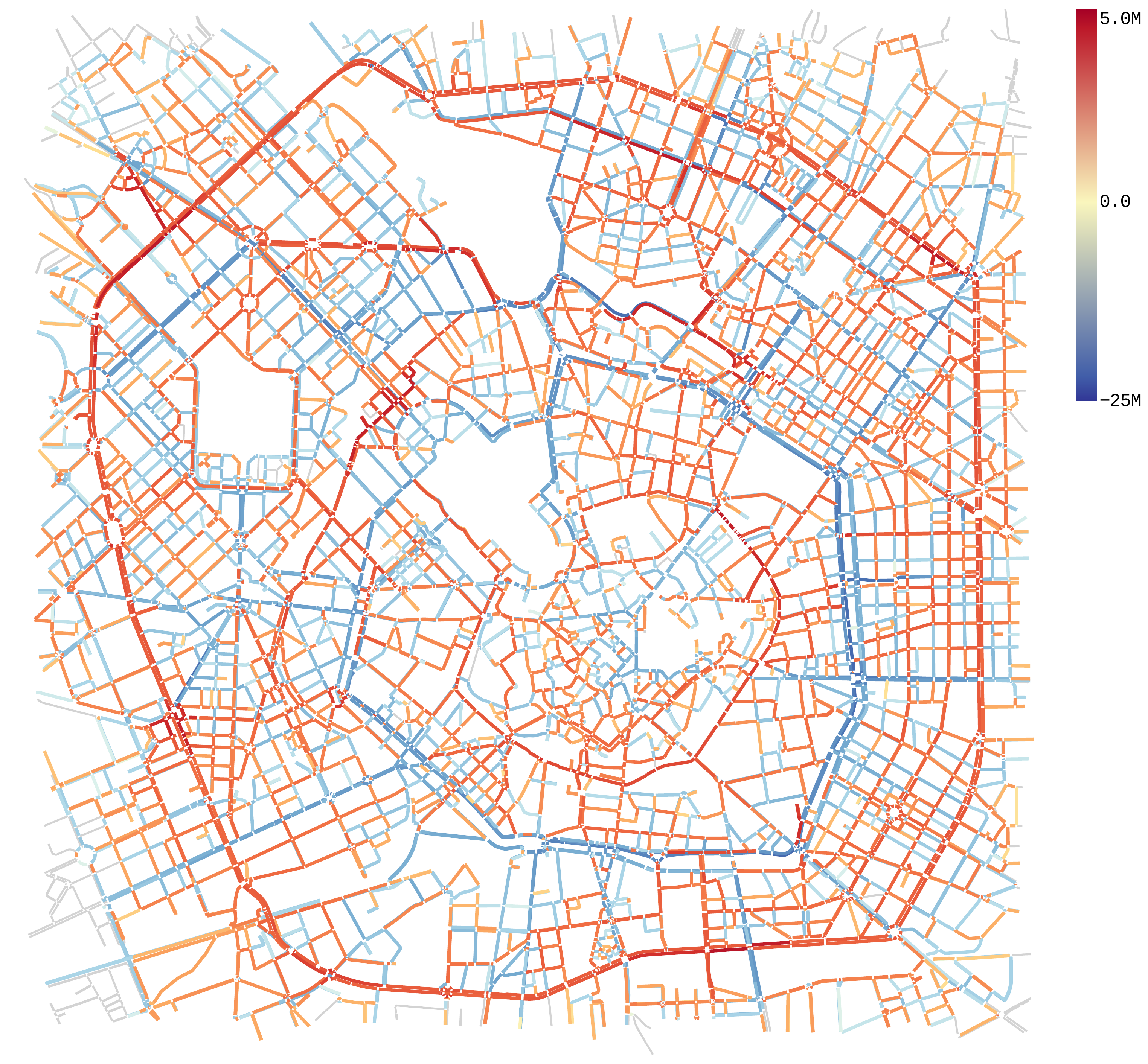}}
    \caption{
    The difference in the total CO$_2$ emitted on each road (in mg per meter of road) when: (a) none of the vehicles is TT-routed and 50\% of them are
    ($\mathcal{E}_0^{\text{\tiny (TT)}}(e) - \mathcal{E}_5^{\text{\tiny (TT)}}(e)$, $\forall e \in E$); (b) all vehicles are TT-routed and 50\% of them are ($\mathcal{E}_{10}^{\text{\tiny (TT)}}(e) - \mathcal{E}_5^{\text{\tiny (TT)}}(e)$, $\forall e \in E$).
    Red roads indicate a positive difference; blue ones indicate a negative one.
    }
    \label{fig:map_diff_TT}
\end{figure}

In Figure \ref{fig:map_diff_TT}a, we show the difference between the per-road emissions (normalized by the road length) when none of the vehicle is TT-routed and when 50\% of them are (i.e., $\mathcal{E}_0^{\text{\tiny (TT)}}(e) - \mathcal{E}_5^{\text{\tiny (TT)}}(e)$, $\forall e \in E$).
Similarly, Figure \ref{fig:map_diff_TT}b shows the normalized emissions difference when all vehicles are TT-routed and 50\% of them are (i.e., $\mathcal{E}_{10}^{\text{\tiny(TT)}}(e) - \mathcal{E}_5^{\text{\tiny(TT)}}(e)$, $\forall e \in E$).
The results are the same as for OSM: when 100\% of vehicles are tt-routed, the emissions are more concentrated towards Milan's ring road (Figure \ref{fig:map_diff_TT}b).
In contrast, when none of them is TT-routed, the emissions are more concentrated towards the city centre (Figure \ref{fig:map_diff_TT}a).

\end{document}